\documentclass[twoside,twocolumn,9pt]{article}
\usepackage{extsizes}
\usepackage[super,sort&compress,comma]{natbib} 
\usepackage[version=3]{mhchem}
\usepackage[left=1.5cm, right=1.5cm, top=1.785cm, bottom=2.0cm]{geometry}
\usepackage{balance}
\usepackage{mathptmx}
\usepackage{sectsty}
\usepackage{graphicx} 
\usepackage{lastpage}
\usepackage[format=plain,justification=justified,singlelinecheck=false,font={stretch=1.125,small,sf},labelfont=bf,labelsep=space]{caption}
\usepackage{float}
\usepackage{fancyhdr}
\usepackage{fnpos}
\usepackage[english]{babel}
\addto{\captionsenglish}{%
  
}
\usepackage{array}
\usepackage{droidsans}
\usepackage{charter}
\usepackage[T1]{fontenc}
\usepackage[usenames,dvipsnames]{xcolor}
\usepackage{setspace}
\usepackage[compact]{titlesec}
\usepackage{hyperref}
\usepackage{bm}
\usepackage{booktabs}
\usepackage{multirow}
\usepackage{algorithm}
\usepackage{algpseudocode}
\usepackage{threeparttable}
\usepackage{cleveref}
  	\crefname{figure}{Figure}{Figures}
  	\crefname{table}{Table}{Tables}
  	\crefname{equation}{Eq.}{Eqs.}
  	\crefname{section}{Section}{Sections}
  	\crefname{subsection}{Section}{Sections}
  	\crefname{subsubsection}{Section}{Sections}
  	\crefname{algorithm}{Algorithm}{Algorithms}
\newcommand{\code}[1]{\texttt{#1}}

\usepackage{todonotes}

\usepackage{epstopdf}%This line makes .eps figures into .pdf - please comment out if not required.

\definecolor{cream}{RGB}{222,217,201}

\begin{document}

\pagestyle{fancy}
\thispagestyle{plain}
\fancypagestyle{plain}{
%%%HEADER%%%
\renewcommand{\headrulewidth}{0pt}
}
%%%END OF HEADER%%%

%%%PAGE SETUP - Please do not change any commands within this section%%%
\makeFNbottom
\makeatletter
\renewcommand\LARGE{\@setfontsize\LARGE{15pt}{17}}
\renewcommand\Large{\@setfontsize\Large{12pt}{14}}
\renewcommand\large{\@setfontsize\large{10pt}{12}}
\renewcommand\footnotesize{\@setfontsize\footnotesize{7pt}{10}}
\makeatother

\renewcommand{\thefootnote}{\fnsymbol{footnote}}
\renewcommand\footnoterule{\vspace*{1pt}% 
\color{cream}\hrule width 3.5in height 0.4pt \color{black}\vspace*{5pt}} 
\setcounter{secnumdepth}{5}

\makeatletter 
\renewcommand\@biblabel[1]{#1}            
\renewcommand\@makefntext[1]% 
{\noindent\makebox[0pt][r]{\@thefnmark\,}#1}
\makeatother 
\renewcommand{\figurename}{\small{Fig.}~}
\sectionfont{\sffamily\Large}
\subsectionfont{\normalsize}
\subsubsectionfont{\bf}
\setstretch{1.125} %In particular, please do not alter this line.
\setlength{\skip\footins}{0.8cm}
\setlength{\footnotesep}{0.25cm}
\setlength{\jot}{10pt}
\titlespacing*{\section}{0pt}{4pt}{4pt}
\titlespacing*{\subsection}{0pt}{15pt}{1pt}
%%%END OF PAGE SETUP%%%

%%%FOOTER%%%
\fancyfoot{}
\fancyfoot[LO,RE]{\vspace{-7.1pt}\includegraphics[height=9pt]{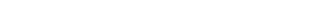}}
\fancyfoot[CO]{\vspace{-7.1pt}\hspace{11.9cm}\includegraphics{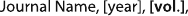}}
\fancyfoot[CE]{\vspace{-7.2pt}\hspace{-13.2cm}\includegraphics{head_foot/RF}}
\fancyfoot[RO]{\footnotesize{\sffamily{1--\pageref{LastPage} ~\textbar  \hspace{2pt}\thepage}}}
\fancyfoot[LE]{\footnotesize{\sffamily{\thepage~\textbar\hspace{4.65cm} 1--\pageref{LastPage}}}}
\fancyhead{}
\renewcommand{\headrulewidth}{0pt} 
\renewcommand{\footrulewidth}{0pt}
\setlength{\arrayrulewidth}{1pt}
\setlength{\columnsep}{6.5mm}
\setlength\bibsep{1pt}
%%%END OF FOOTER%%%

%%%FIGURE SETUP - please do not change any commands within this section%%%
\makeatletter 
\newlength{\figrulesep} 
\setlength{\figrulesep}{0.5\textfloatsep} 

\newcommand{\topfigrule}{\vspace*{-1pt}% 
\noindent{\color{cream}\rule[-\figrulesep]{\columnwidth}{1.5pt}} }

\newcommand{\botfigrule}{\vspace*{-2pt}% 
\noindent{\color{cream}\rule[\figrulesep]{\columnwidth}{1.5pt}} }

\newcommand{\dblfigrule}{\vspace*{-1pt}% 
\noindent{\color{cream}\rule[-\figrulesep]{\textwidth}{1.5pt}} }

\makeatother
%%%END OF FIGURE SETUP%%%

%%%TITLE, AUTHORS AND ABSTRACT%%%
\twocolumn[
  \begin{@twocolumnfalse}
{\includegraphics[height=30pt]{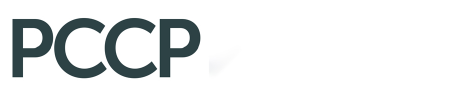}\hfill\raisebox{0pt}[0pt][0pt]{\includegraphics[height=55pt]{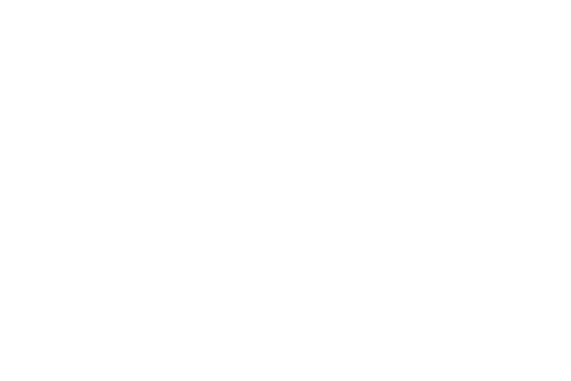}}\\[1ex]
\includegraphics[width=18.5cm]{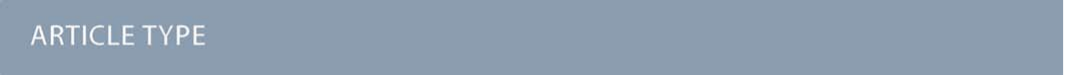}}\par
\vspace{1em}
\sffamily
\begin{tabular}{m{4.5cm} p{13.5cm} }

\includegraphics{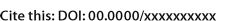} & \noindent\LARGE{\textbf{Economical Quasi-Newton Unitary Optimization of Electronic Orbitals$^\dag$}} \\%Article title goes here instead of the text "This is the title"
\vspace{0.3cm} & \vspace{0.3cm} \\

 & \noindent\large{Samuel A. Slattery, Kshitijkumar A. Surjuse, Charles C. Peterson$^b$, Deborah A. Penchoff$^c$, and Edward F. Valeev\textit{$^{a}$}} \\%Author names go here instead of "Full name", etc.

\includegraphics{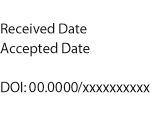} & \noindent\normalsize{We present an efficient quasi-Newton orbital solver optimized to reduce the number of gradient evaluations and other computational steps of comparable cost. The solver optimizes orthogonal orbitals by sequences of unitary rotations generated by the (preconditioned) limited-memory Broyden-Fletcher-Goldfarb-Shanno (L-BFGS) algorithm  equipped with trust-region step restriction. The low-rank structure of the L-BFGS inverse Hessian is exploited when solving the trust-region problem. The efficiency of the proposed ``Quasi-Newton Unitary Optimization with Trust-Region'' (QUOTR) solver is compared to that of the standard Roothaan-Hall approach accelerated by the Direct Inversion of Iterative Subspace (DIIS), and other exact and approximate Newton solvers for mean-field (Hartree-Fock and Kohn-Sham) problems.}

\end{tabular}

 \end{@twocolumnfalse} \vspace{0.6cm}

  ]
%%%END OF TITLE, AUTHORS AND ABSTRACT%%%

%%%FONT SETUP - please do not change any commands within this section
\renewcommand*\rmdefault{bch}\normalfont\upshape
\rmfamily
\section*{}
\vspace{-1cm}

%%%FOOTNOTES%%%

\footnotetext{\textit{$^{a}$~Department of Chemistry, Virginia Tech, Blacksburg, VA 24061, USA. E-mail: efv@vt.edu}}
\footnotetext{\textit{$^{b}$~Office of Advanced Research Computing, University of California, Los Angeles, CA 90095, USA.}}
\footnotetext{\textit{$^{c}$~UT Innovative Computing Laboratory, University of Tennessee, Knoxville, TN 37996, USA.}}

%Please use \dag to cite the ESI in the main text of the article.
%If you article does not have ESI please remove the the \dag symbol from the title and the footnotetext below.
\footnotetext{\dag~Electronic Supplementary Information (ESI) available: geometries (xyz format), convergence statistics and converged energies for most calculations. See DOI: 10.1039/cXCP00000x/}
%additional addresses can be cited as above using the lower-case letters, c, d, e... If all authors are from the same address, no letter is required

%%%END OF FOOTNOTES%%%

%%%MAIN TEXT%%%%
\section{Introduction}
\label{section:introduction}
Orbital optimization is a fundamental ingredient of the electronic structure methods at all levels of approximation, from 1-body models (Hartree-Fock (HF), Kohn-Sham Density Functional Theory (KS DFT), collectively known as the Self-Consistent Field (SCF) method\cite{VRG:hartree:1947:RPP}\footnote{Sometimes SCF is also used to denote a specific class of {\em solvers} to the Hartree-Fock/Kohn-Sham equations, which goes counter to the original use of the term.}), to many-body methods (e.g., multiconfiguration self-consistent field (MCSCF)). Despite the long history of innovation,\cite{VRG:roothaan:1951:RMP, VRG:mcweeny:1956:PRSLA, VRG:levy:1968:IJQC, VRG:hillier:1970:PRSLA, VRG:hillier:1970:IJQC, VRG:levy:1973:CPL, VRG:seeger:1976:JCP, VRG:douady:1980:JCP, VRG:pulay:1980:CPL, VRG:bacskay:1981:CP, VRG:bacskay:1982:CP, VRG:pulay:1982:JCC, VRG:head-gordon:1988:JPC, VRG:fischer:1992:JPC, VRG:shepard:1993:TCA, VRG:rendell:1994:CPL, VRG:wong:1995:JCC, VRG:chaban:1997:TCA, VRG:daniels:2000:PCCP, VRG:vanvoorhis:2002:MP, VRG:vandevondele:2003:JCP, VRG:thogersen:2004:JCP, VRG:thogersen:2005:JCP, VRG:yang:2007:SJSC, VRG:salek:2007:JCP, VRG:weber:2008:JCP, VRG:host:2008:JCP, VRG:baarman:2011:JCP} development of improved orbital optimizers continues to this day.\cite{VRG:sun:2017:AP,VRG:helmich-paris:2021:JCP, VRG:nottoli:2021:MP, VRG:ivanov:2021:CPC, VRG:seidl:2022:JCTC, VRG:dittmer:2023:JCP}
Although the relevant functionals of the orbitals are nonconvex, and global {\em and} local nonconvex optimization is NP-hard,\cite{VRG:vavasis:1991:,VRG:boumal:2019:IJNA} it is known that many practical orbital optimization problems are easily solved using existing heuristics. For the crucial HF/KS SCF use case, the most popular solvers in the molecular context are based on the Roothaan-Hall (RH) iterative diagonalization of the Fock matrix\cite{VRG:roothaan:1951:RMP,VRG:hall:1951:PRSLAa} augmented by convergence accelerators such as Anderson mixing\cite{VRG:anderson:1965:JA} or the closely related direct inversion in the iterative subspace (DIIS) method,\cite{VRG:pulay:1980:CPL, VRG:pulay:1982:JCC, VRG:kudin:2002:JCP, VRG:garza:2012:JCP} as well as others.\cite{VRG:harrison:2004:JCC}
However, several issues plague the efficient RH/DIIS heuristics:
\begin{itemize}
\item for systems with complex electronic structure (such as molecules far from equilibrium, open-shell systems,\cite{VRG:kollmar:1996:JCP,VRG:nottoli:2021:MP} and systems with small HOMO-LUMO gaps\cite{VRG:host:2008:JCP}) convergence will be slow,\cite{VRG:sun:2017:} erratic, or nonexistent,\cite{VRG:helmich-paris:2021:JCP,VRG:rudberg:2012:JPCM}
\item the use of diagonalization produces canonical orbitals whose lack of localization makes them incompatible with fast algorithms for the Fock matrix construction (e.g., using local or sparse density fitting\cite{VRG:koppl:2016:JCTC,VRG:lewis:2016:JCTC,VRG:wang:2020:JCP}),
\item applications to large systems and/or in non-LCAO representations can be
bottlenecked by the $\mathcal{O}(N^3)$ cost of diagonalization,\cite{VRG:rendell:1994:CPL, VRG:wong:1995:JCC,VRG:salek:2007:JCP}
\item locating non-Aufbau (e.g., excited state) solutions is possible\cite{VRG:gilbert:2008:JPCA} but is not robust, and
\item even in favorable cases the convergence rate is 
linear\cite{VRG:stanton:1981:JCP,VRG:fletcher:1970:MP} (i.e., the error is reduced by approximately the same factor each iteration) or perhaps slightly better when accelerated with DIIS\cite{VRG:chupin:2021:EM}; this is slower than the quadratic convergence exhibited by, e.g., the Newton method.\cite{VRG:nocedal:2006:}
\end{itemize}
The lack of convergence guarantees is probably the most severe of these in practice.
Extensions of the standard RH/DIIS heuristics have been devised to improve the robustness\cite{VRG:kollmar:1997:IJQC,VRG:kudin:2002:JCP,VRG:hu:2010:JCP} but for challenging cases the user is expected to control the many heuristic solver control parameters that help the convergence (level shift, damping, etc.).

Orbital optimizer solvers that rely on direct energy minimization can address some/all of these concerns and thus have a long history of development.\cite{VRG:mcweeny:1956:PRSLA,VRG:hillier:1970:PRSLA,VRG:hillier:1970:IJQC,VRG:seeger:1976:JCP,VRG:douady:1980:JCP,VRG:bacskay:1981:CP,VRG:bacskay:1982:CP,VRG:head-gordon:1988:JPC, VRG:fischer:1992:JPC, VRG:shepard:1993:TCA, VRG:rendell:1994:CPL, VRG:wong:1995:JCC, VRG:chaban:1997:TCA, VRG:vanvoorhis:2002:MP,VRG:vandevondele:2003:JCP, VRG:francisco:2004:JCP, VRG:thogersen:2004:JCP, VRG:thogersen:2005:JCP, VRG:yang:2007:SJSC, VRG:salek:2007:JCP, VRG:weber:2008:JCP, VRG:host:2008:JCP, VRG:baarman:2011:JCP}
In the molecular mean-field context direct minimization SCF solvers have long been employed as the recommended alternative in the case of convergence problems, used in combination with RH/DIIS to gain superlinear convergence, and to enable reduced-scaling SCF approaches.\cite{VRG:salek:2007:JCP,VRG:host:2008:JCP}
Nevertheless, RH/DIIS remains the default SCF solver, not due to its formal advantages, but due to its superior efficiency. This may be puzzling since direct minimization solvers are often demonstrated to converge in as few as (or fewer) {\em iterations} than RH/DIIS. \cite{VRG:vanvoorhis:2002:MP,VRG:dittmer:2023:JCP} However, the number of iterations is a misleading figure since each update of the orbitals or density matrix may involve multiple energy/gradient evaluations or solving similarly expensive subproblems (such as multiplication of a trial orbital rotation by the orbital Hessian). In other words, the number of gradient evaluations (Fock build {equivalents}, $N_\mathrm{F}$) in a direct minimization solver is typically significantly greater than the number of iterations ($N_\mathrm{I}$), whereas in RH/DIIS they are equal. Thus the latter typically involves significantly fewer Fock matrix evaluations, which in most practical applications determines the overall cost.

The objective of this work is to design a quasi-Newton orbital optimizer that minimizes the number of gradient evaluations (and its equivalents) to be as competitive with RH/DIIS as possible, and as robust as possible without the need to adjust the control parameters. 
Our ``Quasi-Newton Unitary Optimization with Trust-Region'' (QUOTR) solver uses preconditioned limited-memory Broyden–Fletcher–Goldfarb–Shanno (L-BFGS) algorithm\cite{VRG:nocedal:2006:} step-restricted by trust-region (TR) and leverages the inverse L-BFGS Hessian's low-rank structure to efficiently solve the trust-region update problem.\cite{VRG:burdakov:2017:MPC}

The rest of the manuscript is structured as follows.
In \cref{section:formalism} we briefly review the general classes of SCF solvers before describing the theoretical aspects of QUOTR.
Next, the implementation of QUOTR is discussed in \cref{section:details}. 
In \cref{section:results} we display solver performance statistics for a standard set of chemical systems and make a comparison to a method that uses information from the ``exact'' Hessian. 
Additionally, in \cref{section:results} we illustrate the utility of QUOTR for several prototypical problems where RH/DIIS and other SCF solvers struggle, such as a system with vanishing HOMO-LUMO gap as well as select d- and f-element containing systems.
In \cref{section:conclusion} we summarize our findings.

\section{Formalism}
\label{section:formalism}

\subsection{Overview of SCF Solver Approaches}

All SCF methods attempt to iteratively minimize the electronic energy $E(\mathbf{x})$ or its Lagrangian counterpart, where $\mathbf{x}$ is a set of independent parameters defining the particular method. In practice the minimum is determined by using the energy, its gradient $\mathbf{g}$, and optionally the Hessian {\bf B}. Starting with an initial (guess) set of parameters $\mathbf{x}^{(0)}$ SCF solvers construct improved parameter values using the current energy and its derivatives, (optionally) their values from previous iterations (histories), as well as any optional additional parameters and their histories:
\begin{align}
\label{eq:SCF}
    \mathbf{x}^{(k+1)} = f(\{\mathbf{x}^{(k)}\}, \{E^{(k)}\}, \{\mathbf{g}^{(k)}\}, \dots)
\end{align}
The SCF solvers differ in how they construct the update in \cref{eq:SCF}; unfortunately, it is not possible to systematically classify the solvers since in the vast majority of cases $f()$ is an {\em algorithm}, not a simple function. Thus here we only focus on essential common elements of all SCF solvers.

Most solvers split the update problem \eqref{eq:SCF} into 2 subproblems by defining the parameter update,
\begin{align}
\label{eq:step}
\mathbf{s}^{(k)} \equiv \mathbf{x}^{(k+1)} - \mathbf{x}^{(k)} = \alpha^{(k)} \mathbf{p}^{(k)} 
\end{align}
in terms of a search direction $\mathbf{p}^{(k)}$ and a step size $\alpha^{(k)}$, each of which has its own prescription similar to \cref{eq:SCF}
\begin{align}
\label{eq:SCF-direction}
\mathbf{p}^{(k)} = & g(\{\mathbf{x}^{(k)}\}, \{E^{(k)}\}, \{\mathbf{g}^{(k)}\}, \dots), \\
\label{eq:SCF-stepsize}
\alpha^{(k)} = & h(\{\mathbf{x}^{(k)}\}, \{E^{(k)}\}, \{\mathbf{g}^{(k)}\}, \dots).
\end{align}
The need to control the step size is common to all SCF solvers due to the fundamental nonlinearity of the energy function. Therefore even solvers that do not employ \cref{eq:step}, such as RH/DIIS, still introduce ad hoc ways to control the step size by level shifting, damping, and other means of step restriction.

The simplest ``2-step'' solver is the steepest descent (SD) method\cite{VRG:mcweeny:1956:PRSLA} in which the search direction $\mathbf{p}^{(k)}$ is opposite to the current gradient $\mathbf{g}^{(k)}$:
\begin{align}
\label{eq:SD}
    \mathbf{p}^{(k)} \overset{\mathrm{SD}}{=} & - \frac{\mathbf{g}^{(k)}}{||\mathbf{g}^{(k)}||}
\end{align}
Unfortunately, although the SD method is guaranteed to converge to a nearby minimum, the plain SD variant converges very slowly;\cite{VRG:claxton:1971:TCA,VRG:sleeman:1968:TCA} this can be rationalized by comparing it to the (exact) Newton step:
\begin{align}
\mathbf{s}^{(k)} \overset{\mathrm{Newton}}{=} & - \left(\mathbf{B}^{(k)}\right)^{-1} \mathbf{g}^{(k)}.
\end{align}
Hessian $\mathbf{B}$ is a diagonally-dominant matrix with 
a large (and growing with the basis set size) condition number. Luckily it is relatively simple to construct an effective approximation to the Hessian; a particularly popular way is to use only the 1-electron terms in the Hessian, $\mathbf{B}_{\mathrm{1e}}$.
Approximate Hessians can then be used for preconditioning SD (using the 1-electron Hessian for preconditioning is also known as the ``energy weighted steepest descent'' method \cite{VRG:hillier:1970:PRSLA,VRG:hillier:1970:IJQC,VRG:seeger:1976:JCP})
by replacing $\mathbf{g}^{(k)}$ in \cref{eq:SD} with the preconditioned gradient:
\begin{align}
\tilde{\mathbf{g}}^{(k)} \equiv \left(\mathbf{B}_{\mathrm{1e}}^{(k)}\right)^{-1} \mathbf{g}^{(k)}.
\end{align}
The RH method can be viewed as a simplified version of preconditioned SD, due to its step being exactly the negative of the gradient preconditioned by the 1-electron Hessian:\cite{VRG:thogersen:2005:JCP,VRG:host:2008:JCP}
\begin{align}
\mathbf{s}^{(k)} \overset{\mathrm{RH}}{=} & - \tilde{\mathbf{g}}^{(k)}.
\end{align}

More sophisticated prescriptions for direction include the conjugate gradient (CG) method\cite{VRG:wong:1995:JCC, VRG:vandevondele:2003:JCP, VRG:daniels:2000:PCCP, VRG:weber:2008:JCP, VRG:baarman:2011:JCP} in which history is limited to the information about the current and previous iteration. Of course, the use of preconditioning is mandatory with CG just as with SD. Unfortunately neither SD nor CG, even with an approximate preconditioner, lead to an optimal convergence rate near the minimum. Thus the most efficient solvers utilize exact or approximate Hessians near the minimum. The time-determining step of such models usually involves direct evaluation of the action of exact (or approximated) Hessian onto a trial step, at a cost similar to the cost of the gradient evaluation (i.e., the Fock matrix evaluation in the mean-field case). \cite{VRG:pulay:1982:JCC, VRG:shepard:1993:TCA, VRG:rendell:1994:CPL} Although it is possible to apply the straightforward Newton method using the exact Hessian when sufficiently close to the minimum,\cite{VRG:douady:1980:JCP} to be able to use the exact Hessian further away from the minimum requires some form of step restriction.
The popular augmented Hessian (AH)\cite{VRG:helmich-paris:2021:JCP} method can be viewed as a Newton method with optimally restricted steps; it can also be viewed as a quasi-Newton method in which an approximate (level-shifted) Hessian is used. The diverse family of quasi-Newton methods each use approximate Hessians of some form, often generated from information contained in the gradients and steps of the previous iterations. \cite{VRG:nocedal:2006:}
The quasi-Newton idea has been used in MCSCF for a long time,\cite{VRG:olsen:1982:JCPa} and the most commonly employed approximation in SCF is some form of the BFGS algorithm. \cite{VRG:head-gordon:1988:JPC, VRG:fischer:1992:JPC, VRG:chaban:1997:TCA, VRG:vanvoorhis:2002:MP, VRG:baarman:2011:JCP, VRG:ivanov:2021:CPC}
The BFGS method has recently been used with success in the MCSCF context \cite{VRG:kreplin:2019:JCP} and the selected configuration interaction (CI) context. \cite{VRG:yao:2021:JCTC}

Although some solvers compute the step length separately from the direction, more sophisticated approaches fuse step restriction deeper into the step computation. 
Indeed, when an underlying quadratic model of the energy exists, it is not natural to simply perform a line search toward the (unrestricted) minimum of the model, considering that the model is known to be locally accurate in all directions.
The alternative concept of searching for the minimum of a model in all directions, but restricting the step size to some maximum value, is the key idea of the trust-region (TR) method. \cite{VRG:jorgensen:1983:JCP, VRG:jensen:1984:JCP, VRG:francisco:2004:JCP, VRG:thogersen:2004:JCP, VRG:thogersen:2005:JCP, VRG:yang:2007:SJSC, VRG:salek:2007:JCP, VRG:host:2008:JCP, VRG:helmich-paris:2021:JCP}
Two important aspects of any TR method are: how the trust-region is updated between iterations, and how the trust-region problem is solved for the step.
The update method that is commonly used is based on an algorithm developed by Fletcher, \cite{VRG:fletcher:1980:} and one of the first true TR applications in quantum chemistry used it in the context of MCSCF.\cite{VRG:jorgensen:1983:JCP}
A common occurrence of the TR problem in quantum chemistry is within the framework of the AH method; due to the use of full (level-shifted) Hessian in AH the cost of the TR solve is similar to the cost of the unrestricted step.\cite{VRG:helmich-paris:2021:JCP}
Here we use the TR method in the context of the L-BFGS method which allows us to exploit the low-rank structure of the L-BFGS Hessian to essentially eliminate the extra cost of using the TR method.\cite{VRG:burdakov:2017:MPC}

\subsection{QUOTR: Quasi-Newton Unitary Optimization with Trust-Region}

Our direct minimization SCF solver is a preconditioned quasi-Newton (L-BFGS) solver with TR step restriction. Although its aspects are similar to prior SCF solvers, there are several novel elements:

\begin{itemize}
\item The optimization is parameterized with a consistent ``reference'' (epoch) MO basis allowing use of the exact gradient with minimal computation after the Fock matrix is constructed.
\item The preconditioner is updated only on some iterations, and it is regularized in a simple way to ensure a positive definite Hessian.
\item The low-rank structure of the L-BFGS Hessian is exploited when solving for the quasi-Newton step on the TR boundary.
\end{itemize}

The QUOTR algorithm is described in \cref{algorithm:scf_solver}, and its user-controllable parameters are listed in \cref{table:scf_parameters}.
The parameters listed have been divided into three groups: free user choice, convergence tweaking, and expert-only controls.
The two convergence criteria for the solver can be chosen however the user wishes, within reason.
The next three parameters could be adjusted in cases that convergence is not as fast as desired.
Finally, the remaining parameters are not recommended to be adjusted.
Below we elaborate on each key aspect of the solver.

\begin{algorithm*}
\caption{QUOTR SCF Solver}
\begin{algorithmic}[1]

\Function{QUOTR}{$\mathbf{C}^{(0)}$}
\State $k \gets 0$, $S_{\mathrm{taken}} \gets$ \code{false}, $R_{\mathrm{hist}} \gets $ \code{false}, $\mathbf{U}_\mathrm{epoch} \gets \mathbf{1}$, $\{E^{(k)}, \mathbf{F}_\mathrm{AO}\} \gets \Call{Fock}{\mathbf{C}^{(k)}}$
\State $\mathbf{g}^{(k)} \gets \Call{Grad}{\mathbf{F}_\mathrm{AO}, \mathbf{C}^{(k)}, \mathbf{U}_\mathrm{epoch}}$ \Comment{\cref{eq:gradient_epoch}}
\State $\mathbf{B}_0 \gets \Call{Preconditioner}{\mathbf{F}_\mathrm{AO}, \mathbf{C}^{(k)}}$ \Comment{\cref{eq:diag_hessian}}
\While{$\mathrm{RMS}(\mathbf{g}^{(k)}) > t_{cg}$ AND (k=0 OR $\Delta E^{(k)} > t_{ce}$)}

\If{$S_{\mathrm{taken}}$}
\State $\mathbf{s}^{(k-1)} \gets \mathbf{B}_0^{-1/2} \tilde{\mathbf{s}}^{(k-1)}$
\If{$\mathbf{y}^{(k-1)}\cdot\mathbf{s}^{(k-1)} > t_r ||\mathbf{y}^{(k-1)}|| \, || \mathbf{s}^{(k-1)}||$}
\State $\tilde{\mathbf{y}}^{(k-1)} \gets \mathbf{B}_0^{-1/2} \mathbf{y}^{(k)}$
\State $\mathbf{\tilde{S}} \gets \Call{Append}{\mathbf{\tilde{S}}, \mathbf{\tilde{s}}^{(k-1)}}$, $\mathbf{\tilde{S}} \gets \Call{Trim}{\mathbf{\tilde{S}},m}$, $\tilde{\mathbf{Y}} \gets \Call{Append}{\mathbf{\tilde{Y}}, \mathbf{\tilde{y}}^{(k-1)}}$, $\mathbf{\tilde{Y}} \gets \Call{Trim}{\mathbf{\tilde{Y}},m}$ $\tilde{\mathbf{V}} \gets \Call{Concat}{\tilde{\mathbf{S}},\tilde{\mathbf{Y}}}$ \Comment{update history}
\EndIf
\EndIf

\State $R_{\mathrm{hist}} \gets$ $\Delta^{(k)} < t_t$ OR $||\mathbf{g}^{(k)}||_{\infty} > t_b$
\If{NOT $R_{\mathrm{hist}}$}
\State $\tilde{\mathbf{g}}^{(k)} \gets \mathbf{B}_0^{-1/2} \mathbf{g}^{(k)}$
\If{$k > 0$ AND $\tilde{\mathbf{V}}$ not empty} \Comment{BFGS}
\State $\tilde{\mathbf{s}}^{(k)}$ $\gets$ \Call{L-BFGS}{$\tilde{\mathbf{g}}^{(k)}, \mathbf{\tilde{V}}$} \Comment{\cref{eq:newton-step-preconditioned,eq:H-BFGS-preconditioned}}
\If {$||\tilde{\mathbf{s}}^{(k)}|| > \Delta^{(k)}$}
\State $\tilde{\mathbf{s}}^{(k)} \gets \Call{TRStep}{\Delta^{(k)},\tilde{\mathbf{g}}^{(k)},\tilde{\mathbf{s}}^{(k)},\tilde{\mathbf{V}}}$ \Comment{\cref{algorithm:tr_solver}}
\EndIf
\State $q^{(k)} \gets $ \cref{eq:qk}, $R_{\mathrm{hist}} \gets$ $q^{(k)} > 0$ \Comment{energy increase predicted}
\Else
\State $\tilde{\mathbf{s}}^{(k)} \gets -\tilde{\mathbf{g}}^{(k)}$ \Comment{steepest descent}
\EndIf
\EndIf

\If{$R_{\mathrm{hist}}$} \Comment{new epoch}
\If{not $S_\mathrm{taken}$}
\State $\{E^{(k)}, \mathbf{F}_\mathrm{AO}\} \gets \Call{Fock}{\mathbf{C}^{(k)}}$
\EndIf
\State $\tilde{\mathbf{V}} \gets \{\}$,  $\mathbf{U}_{\mathrm{epoch}} \gets \mathbf{1}$, $\mathbf{B}_0 \gets \Call{Preconditioner}{\mathbf{F}_\mathrm{AO}, \mathbf{C}^{(k)}}$, $\mathbf{g}^{(k)} \gets \Call{Grad}{\mathbf{F}_\mathrm{AO}, \mathbf{C}^{(k)}, \mathbf{U}_\mathrm{epoch}}$
\State $\tilde{\mathbf{g}}^{(k)} \gets \mathbf{B}_0^{-1/2} \mathbf{g}^{(k)}$, $\tilde{\mathbf{s}}^{(k)} \gets -\tilde{\mathbf{g}}^{(k)}$ \Comment{steepest descent}
\EndIf

\If{$k = 0$ OR $\tilde{\mathbf{V}}$ is empty}   \Comment{line search} 
\State $\alpha^{(k)} \gets \Call{LineSearch}{\tilde{\mathbf{s}}^{(k)}}$ \Comment{\cref{sec:line-search}}
\State $\tilde{\mathbf{s}}^{(k)} \gets \alpha^{(k)} \tilde{\mathbf{s}}^{(k)}/||\tilde{\mathbf{s}}^{(k)}||$, $\Delta^{(k)} \gets \alpha^{(k)}$
\EndIf

\State $S_\mathrm{taken} \gets \code{true}, \mathbf{s}^{(k)} \gets \mathbf{B}^{-1/2}_0 \tilde{\mathbf{s}}^{(k)}$ \Comment{attempt step}
\State $\mathbf{U}^{(k)} \gets \Call{UnitaryStep}{\mathbf{s}^{(k)}, \mathbf{U}_\mathrm{epoch}}$ \Comment{\cref{eq:sigma_from_epoch},\cref{eq:unitary_step}}
\State $\mathbf{C}^{(k+1)} \gets \mathbf{C}^{(k)} \mathbf{U}^{(k)}$, $\{E^{(k+1)}, \mathbf{F}_\mathrm{AO}\} \gets \Call{Fock}{\mathbf{C}^{(k+1)}}$, $\Delta E^{(k)} \gets E^{(k+1)} - E^{(k)}$ 

\If{$k > 0$ AND $\tilde{\mathbf{V}}$ not empty}
\State $\rho^{(k)} \gets \Delta E^{(k)}/q^{(k)}$
\State $S_{\mathrm{taken}} \gets $ $\rho^{(k)} \geq \tau_1$ OR $\Delta E^{(k)} \leq t_0$
\If{$\rho^{(k)} < \tau_2$}
\State $\Delta^{(k+1)} \gets \Call{min}{\eta_1 \Delta^{(k)}, \eta_2 ||\tilde{\mathbf{s}}^{(k)}||}$
\ElsIf{$(\rho^{(k)} > \tau_3$ AND $||\tilde{\mathbf{s}}^{(k)}|| > \eta_3\Delta^{(k)})$}
\State $\Delta^{(k+1)} \gets \eta_4 \Delta^{(k)}$
\EndIf
\Else
\State $S_{\mathrm{taken}} \gets$ $\Delta E^{(k)} \leq t_0$
\If{$S_{\mathrm{taken}}$}
\State $\Delta^{(k+1)} \gets ||\tilde{\mathbf{s}}^{(k)}||$
\EndIf
\EndIf

\If{$S_{\mathrm{taken}}$} 
\State $k \gets k + 1$,  $\mathbf{U}_\mathrm{epoch} \gets \mathbf{U}_\mathrm{epoch} \mathbf{U}^{(k-1)}$, $\mathbf{g}^{(k)} \gets \Call{Grad}{\mathbf{F}_\mathrm{AO}, \mathbf{C}^{(k)}, \mathbf{U}_\mathrm{epoch}}$, $\mathbf{y}^{(k-1)} \gets \mathbf{g}^{(k)} - \mathbf{g}^{(k-1)}$ \Comment{accept step}
\Else
\State $\Delta^{(k)} \gets \Delta^{(k+1)}$ \Comment{reject step, change TR}
\EndIf
\EndWhile

\State \Return $\mathbf{C}^{(k)}$
\EndFunction

\end{algorithmic}
\label{algorithm:scf_solver}
\end{algorithm*}

\begin{table}[ht]
    \centering
    \caption{User-controllable parameters of the QUOTR solver.}
    \label{table:scf_parameters}
    \begin{tabular*}{0.47\textwidth}{@{\extracolsep{\fill}}lll}
        \hline
        Description & Symbol & Value \\
        \hline
        energy convergence threshold & $t_{ce}$ & $10^{-9}$ \\
        gradient convergence threshold & $t_{cg}$ & $10^{-5}$ \\
        \hline
        L-BFGS start threshold & $t_b$ & 0.1 \\
        max history size & $m$ & 8 \\
        regularizer threshold & $t_{r}$ & 0.25 \\
        \hline
        history keep threshold & $t_h$ & $10^{-5}$ \\
        exponential tolerance & $t_e$ & $10^{-15}$ \\
        Compare zero threshold & $t_0$ & $10^{-11}$ \\
        line search fitting range shrink factor & $\alpha_\mathrm{fit,shrink}$ & 1/2 \\
        minimum TR tolerance & $t_t$ & $10^{-10}$ \\
        TR step accept threshold & $\tau_1$ & 0 \\
        TR shrink threshold & $\tau_2$ & 0.25 \\
        TR expand threshold & $\tau_3$ & 0.75 \\
        TR shrink factor & $\eta_1$ & 0.25 \\
        TR shrink by step factor & $\eta_2$ & 0.5 \\
        TR expand check factor & $\eta_3$ & 0.8 \\
        TR expand by factor & $\eta_4$ & 2.0 \\
        \hline
    \end{tabular*}
\end{table}

\subsubsection{Parameterization.}

It is important to consider how the standard unconstrained quasi-Newton minimization scheme can be mapped to the constrained minimization of the single-determinant energy where the orbitals are required to be orthonormal. 
In the following, we assume the linear combination of atomic orbitals (LCAO) representation of the molecular orbitals (MOs), and thus the MOs are defined by the coefficient matrix $\mathbf{C}$.

We seek a unitary matrix, $\bar{\mathbf{U}}$, that transforms the initial (guess) set of orthonormal orbitals, $\mathbf{C}^{(0)}$, into the target solution, $\bar{\mathbf{C}}$.
\begin{align}
\bar{\mathbf{C}} = \mathbf{C}^{(0)} \bar{\mathbf{U}}
\label{eq:solver_goal}
\end{align}
The target unitary matrix is built as a sequence of unitary rotations,
\begin{align}
\bar{\mathbf{U}} = \mathbf{U}^{(0)} \mathbf{U}^{(1)} \dots \equiv \prod_{k} \mathbf{U}^{(k)},
\label{eq:u_total}
\end{align}
with each unitary rotation $\mathbf{U}^{(k)}$ obtained by solving a local subproblem.
The orbitals at iteration $k > 0$ are given by the coefficient matrix obtained from the total rotation determined thus far.
\begin{align}
\mathbf{C}^{(k)} = \mathbf{C}^{(0)} \prod_{i=0}^{k-1}\mathbf{U}^{(i)} = \mathbf{C}^{(k-1)} \mathbf{U}^{(k-1)}
\label{eq:ck_first_def}
\end{align}

We use the standard\cite{VRG:thouless:1960:NP, VRG:levy:1969:CPL, VRG:douady:1980:JCP, VRG:helgaker:2000:} exponential parameterization of unitary $\mathbf{U}^{(k)}$ : \begin{align}
\mathbf{U}^{(k)} = \mathrm{exp}(\bm{\sigma}^{(k)}),
\label{eq:unitary_step}
\end{align}
where $\bm{\sigma}^{(k)}$ is an antihermitian matrix encoding the unitary rotation of the orbitals.
Formulating the optimization problem in terms of the antihermitian coordinate matrix $\bm{\sigma}$ rather than in terms of the density matrix allows us to avoid the need for diagonalization (which restores the idempotency of the density matrix in extrapolation/interpolation methods like DIIS\cite{VRG:pulay:1982:JCC,VRG:kudin:2002:JCP}).
The matrix exponentials are evaluated accurately (to finite precision $t_e$) as a Taylor series expansion using a simple scaling-and-squaring approach\cite{VRG:moler:1978:SR, VRG:helmich-paris:2021:JCP} with a fixed order of 2. In this technique, the matrix to be exponentiated is first divided by $2^2 = 4$. Next, the Taylor expansion for the exponential function is carried out on this scaled matrix, truncating when the norm of the next term in the series drops below $t_e$. The resulting matrix is raised to the fourth power (by squaring twice) to obtain the exponential of the original matrix. Although some approaches evaluate the exponential approximately and return to the target manifold by additional orthogonalization,\cite{VRG:douady:1980:JCP,VRG:chaban:1997:TCA}
accurate evaluation is important to be able to maintain the fidelity of the relationship between parameters and the objective function values in the context of extrapolation methods like BFGS.
Evaluation of the matrix exponential could be improved further, e.g., by leveraging the block-sparse antihermitian structure of $\bm{\sigma}$. \cite{VRG:seidl:2022:JCTC}

It is important to express both the gradients and parameter updates for each iteration in the same coordinate frame (basis). In the context of the DIIS methods this is usually done by working in the AO basis (e.g., as noted by Pulay in Ref. \citenum{VRG:pulay:1982:JCC} ``In order to be useful for extrapolation purposes, [the Fock] matrices (one in each iteration step) must be transformed to a common basis, e.g., to the original AO basis set.''). Here the working frame is defined by the initial orthogonal MO basis for each {\em epoch} (namely, the sequence of iterations whose history is used to construct the current approximation to the Hessian). 
The gradient of the energy evaluated with orbitals $\mathbf{C}^{(k)}$ with respect to their arbitrary rotation $\bm{\sigma}^{(k)}$ has the familiar form when expressed in terms of $\mathbf{C}^{(k)}$:
\begin{align}
(\mathbf{g}^{(k)})_{ia} \equiv \frac{\partial E}{\partial \sigma^{(k)}_{ai}} = 2 n_i (\mathbf{F}^{(k)})_{ia},
\label{eq:gradient_current}
\end{align} 
where $a$ and $i$ refer to the unoccupied and occupied MOs in $\mathbf{C}^{(k)}$, respectively, and
$n_i$ is the occupancy of $i$th orbital (2 for spin-restricted closed-shell SCF, 1 for spin-unrestricted SCF).
However, the gradient ``at'' arbitrary MOs can be expressed in an arbitrary (e.g., epoch) basis. For example, the gradient at current orbitals $\mathbf{C}^{(k)}$ in epoch basis can be obtained by transforming \cref{eq:gradient_current} to the epoch basis:
\begin{align}
\mathbf{g}^{(k)}_\mathrm{epoch}\equiv 2 n_i [\mathbf{F}_{\mathrm{epoch}}^{(k)} , \mathbf{P}_{\mathrm{epoch}}^{(k)} ].
\label{eq:gradient_epoch}
\end{align}
Here $\mathbf{F}_{\mathrm{epoch}}^{(k)}$ is the Fock matrix evaluated with current orbitals $\mathbf{C}^{(k)}$ but represented in the epoch MO basis and $\mathbf{P}_{\mathrm{epoch}}^{(k)}$ is the projector onto the occupied MOs in $\mathbf{C}^{(k)}$ expressed in the epoch basis.
In practice, the Fock matrix in the epoch basis is evaluated in AO basis using the AO density matrix evaluated from $\mathbf{C}^{(k)}$ and then transformed to the epoch basis.
The projection operator onto the occupied space at iteration $k$ in the epoch basis is obtained as
\begin{align}
\mathbf{P}^{(k)}_{\mathrm{epoch}} = \mathbf{U}_{\mathrm{epoch}}^{(k)} \mathbf{P} (\mathbf{U}_{\mathrm{epoch}}^{(k)})^{\mathrm{T}},
\label{eq:p_epoch}
\end{align}
where $\mathbf{P}$ is the diagonal matrix, with ones and zeroes on the diagonal for the occupied/unoccupied MOs, respectively.
Not only does this formulation of the gradient allow us to have a consistent basis for forming the L-BFGS Hessian, but it also avoids evaluating the gradient at particular MOs using a non-truncating Taylor series around the reference/epoch basis, as in other solvers.\cite{VRG:ivanov:2021:CPC}

Note that $\bm{\sigma}^{(k)}$ is a matrix, but only some of its elements can be varied independently. It is also traditional in applied mathematics literature to arrange the parameters of  multivariate functions into vectors.
Thus it is appropriate to comment on the detailed relationship of $\bm{\sigma}^{(k)}$ and the corresponding {\em step} vector $\mathbf{s}^{(k)}$.
Due to the antihermiticity of $\bm{\sigma}$, $\sigma_{pq} = -\sigma_{qp}$, hence for real orbitals (which we assume here without any loss of generality) only the lower elements are independent.
Notice that the gradient matrix in \cref{eq:gradient_epoch} is also antihermitian, with the same structure as $\bm{\sigma}$.
Although for 1-body SCF methods considered here elements of the gradient matrix are nonzero only between occupied and unoccupied orbitals, hence one would think that only $(\bm{\sigma})_{ia}$ elements need to be independently varied, this is only true for $\bm{\sigma}$ expressed in the current MO basis (i.e., the basis defining the density). To be able to work in an arbitrary (e.g., epoch) basis all lower-triangular elements of $\bm{\sigma}$ thus must be considered independent.
For an MO basis with $n_o$ orbitals, this means that the number of independent parameters is $n \equiv n_o (n_o -1)/2$, assuming no additional symmetries are taken into account.
In the spin-unrestricted case, the kappa elements for the separate alpha and beta spin MOs are simply concatenated into one vector.
Notice that the gradient matrix in \cref{eq:gradient_epoch} is also antihermitian, with the same structure as $\bm{\sigma}$.
Thus, we can map the matrix elements of the gradient to a vector in exactly the same way as for the matrix elements of $\bm{\sigma}$, using only the lower (or upper) triangle. 
We use the symbol $\mathbf{s}^{(k)}$ for the vector version of $\bm{\sigma}^{(k)}$; for the gradient henceforth only its vector form, $\mathbf{g}^{(k)}$, is used.

The steps and gradient differences are the ingredients for the L-BFGS update to the Hessian and its inverse.
All quantities throughout each epoch (see below) are kept in the epoch ``reference'' basis to make application of L-BFGS and TR consistent.
However, when a new step $\mathbf{s}^{(k)}$ is proposed, to convert it to the unitary rotation via \cref{eq:unitary_step} it must be transformed to the current MO basis, via
\begin{align}
\bm{\sigma}^{(k)}=\mathbf{U}_{\mathrm{epoch}}^{(k-1)\mathrm{T}} \bm{\sigma}_{\mathrm{epoch}}^{(k)} \mathbf{U}_{\mathrm{epoch}}^{(k-1)}.
\label{eq:sigma_from_epoch}
\end{align}

Our convergence criteria require both the energy change between iterations ($\Delta E^{(k)} \equiv E^{(k+1)} - E^{(k)}$) and the root mean square (RMS) of the unique gradient elements to be small.
Generally, we require energy change to be below $1 \times 10^{-9} E_{\mathrm{h}}$ and the RMS gradient (in epoch basis) to be smaller than $1 \times 10^{-5}$.
However, for some comparisons with other solvers, we use the 2-norm of the gradient instead of the RMS version.

\subsubsection{Quasi-Newton method.}
\label{subsection:quasi_Newton}

The L-BFGS algorithm is used to approximate the Hessian, $\mathbf{B}_{\mathrm{BFGS}}^{(k)} \approx \mathbf{B}^{(k)}$, and its inverse, $\mathbf{H}_{\mathrm{BFGS}}^{(k)} \equiv (\mathbf{B}_{\mathrm{BFGS}}^{(k)})^{-1} \approx (\mathbf{B}^{(k)})^{-1}$, using the history vectors from (at most) $m$ previous iterations.
In the following, we will assume that the history size is equal to $m$ for simplicity.
This approximate Hessian and its inverse can be represented in low-rank form as follows:\cite{VRG:burdakov:2017:MPC}
\begin{align}
\mathbf{B}_{\mathrm{BFGS}}^{(k)} = \mathbf{B}_{0} - \mathbf{V}_{\mathrm{B}}^{(k)} (\mathbf{W}^{(k)})^{-1} \mathbf{V}_{\mathrm{B}}^{(k)\mathrm{T}}
\label{eq:hessian_lbfgs}
\end{align}
\begin{align}
\mathbf{H}_{\mathrm{BFGS}}^{(k)} = \mathbf{H}_{0} + \mathbf{V}_{\mathrm{H}}^{(k)} \mathbf{M}^{(k)} \mathbf{V}_{\mathrm{H}}^{(k)\mathrm{T}}
\label{eq:inv_hessian_lbfgs}
\end{align}
Here the matrix $\mathbf{B}_{0}$ is an initial (often diagonal) approximation to the Hessian chosen at the beginning of the current epoch, which in principle could be any positive definite matrix. \cite{VRG:nocedal:2006:}
The matrix $\mathbf{H}_{0}$ is the corresponding initial inverse Hessian approximation: $\mathbf{H}_{0} \equiv \mathbf{B}_{0}^{-1}$.
More will be said about these critical components later.
At iteration $k$ with the BFGS history containing $m$ \{step,gradient differences\} pairs, matrix $\mathbf{V}_{\mathrm{B}}^{(k)}$ has $2m$ columns, with the first $m$ being history step vectors multiplied by the initial Hessian, $\mathbf{B}_{0} \mathbf{s}^{(k)}$, followed by the $m$ matching gradient differences, $\mathbf{y}^{(k)} \equiv \mathbf{g}^{(k+1)} - \mathbf{g}^{(k)}$.
$\mathbf{V}_{\mathrm{H}}^{(k)}$ is obtained from $\mathbf{V}_{\mathrm{B}}^{(k)}$ as $\mathbf{V}_{\mathrm{H}}^{(k)} \equiv \mathbf{H}_{0}\mathbf{V}_{\mathrm{B}}^{(k)}$.
The square matrices $\mathbf{W}^{(k)}$ and $\mathbf{M}^{(k)}$ have dimension $2m$, and the need for the inverse of $\mathbf{W}^{(k)}$ is not a problem since $m$ is typically small (we have used $m = 8$).
The formation of $\mathbf{W}^{(k)}$ and $\mathbf{M}^{(k)}$ is described in the literature,\cite{VRG:byrd:1994:MPa,VRG:burdakov:2017:MPC} but essentially they are composed of various dot products involving the history vectors, (requiring an inverse of one of the $m \times m$ subblock).
\begin{align}
\mathbf{W}^{(k)} = 
\left ( \begin{array}{c|c}
((\mathbf{S}^{(k)})^{\mathrm{T}} \mathbf{B}_{0} \mathbf{S}^{(k)}) &  \mathbf{L}^{(k)} \\
\midrule
   (\mathbf{L}^{(k)})^{\mathrm{T}} & -\mathbf{E}^{(k)}
\end{array}
\right )
\label{eq:w_lbfgs}
\end{align}
\begin{align}
\mathbf{M}^{(k)} = 
\left ( \begin{array}{c|c}
((\mathbf{L}^{(k)})^{-1} (\mathbf{E}^{(k)} + (\mathbf{Y}^{(k)})^{\mathrm{T}} \mathbf{H}_{0} \mathbf{Y}^{(k)}) (\mathbf{L}^{(k)})^{\mathrm{-T}}) & -(\mathbf{L}^{(k)})^{-1} \\
\midrule
   -(\mathbf{L}^{(k)})^{-\mathrm{T}} & \mathbf{0}
\end{array}
\right )
\label{eq:m_lbfgs}
\end{align}

In \cref{eq:w_lbfgs} and \cref{eq:m_lbfgs}, $\mathbf{S}^{(k)}$ is an $n \times m$ matrix containing the history column vectors, $\mathbf{s}^{(k)}$, and similarly $\mathbf{Y}^{(k)}$ contains the gradient differences, $\mathbf{y}^{(k)}$.
The smaller $m \times m$ submatrices $\mathbf{L}^{(k)}$ and $\mathbf{E}^{(k)}$ are simply constructed as below.
\begin{align}
\mathbf{L}_{ij}^{(k)} = \begin{cases}
         \mathbf{s}^{(k-m-1+i)} \cdot \mathbf{y}^{(k-m-1+j)}, & \mathrm{if} \quad i > j \\
         0 & \mathrm{otherwise}
         \end{cases}
\end{align}
\begin{align}
\mathbf{E}_{ij}^{(k)} = \begin{cases}
         \mathbf{s}^{(k-m-1+i)} \cdot \mathbf{y}^{(k-m-1+j)}, & \mathrm{if} \quad i = j \\
         0 & \mathrm{otherwise}
         \end{cases}
\end{align}

One of the advantages of the L-BFGS Hessian approximation, apart from not requiring calculation of second derivatives, is that it can be stored in this factorized form by simply keeping the relatively small matrices $\mathbf{B}_{0}$, $\mathbf{V}_{\mathrm{B}}^{(k)}$, and $\mathbf{W}^{(k)}$.
Considering that $\mathbf{B}_{0}$ is a diagonal matrix, we only need to store $(2m+1)n + 4m^2$ elements, which is typically much smaller than the full Hessian which requires $n^2$ elements.
From the development up to this point, it would seem that we also need to store the information for the inverse L-BFGS Hessian, specifically $\mathbf{V}_{\mathrm{H}}^{(k)}$, but this will be dealt with soon.

The quasi-Newton step, $\mathbf{s}^{(k)}=-\mathbf{H}_{\mathrm{BFGS}}^{(k)}\mathbf{g}^{(k)}$, is calculated by multiplying the inverse L-BFGS Hessian of \cref{eq:inv_hessian_lbfgs} with the negative of the gradient.
Thus, the factorized form of \cref{eq:inv_hessian_lbfgs} makes the task of calculating the quasi-Newton step simply a matter of a few matrix-vector multiplications.
Note that although we only need the inverse Hessian to compute the quasi-Newton step, the Hessian is used to compute the energy decrease predicted by the quadratic model.
This is needed for determining how the TR is to be updated between iterations as described in \cref{subsection:trust_region}.

As is well known, due to the large (and increasing with the basis) condition number of the Hessian it is important to use a preconditioner to achieve competitive convergence.\cite{VRG:wong:1995:JCC,VRG:vanvoorhis:2002:MP,VRG:chaban:1997:TCA}
Since the orbital Hessian is often diagonally dominant and its 1-electron (Fock) contributions are cheap to evaluate, we define $\mathbf{B}_{0}$ in terms of the diagonal elements of the 1-electron component of the exact Hessian, $\mathbf{B}_\mathrm{1e}$,
whose unique nonzero elements in the current MO basis are
\begin{align}
(\mathbf{B}_\mathrm{1e})_{(ia)(jb)} = 2 n_i (F_{ab} \delta_{ij} - F_{ij} \delta_{ab}).
\label{eq:1e_hessian}
\end{align}
Since we compute the preconditioner once per epoch (i.e., infrequently), for each epoch we choose the basis to make the Hessian as diagonally dominant as possible by choosing the ``pseudocanonical'' basis, \cite{VRG:hillier:1970:PRSLA} which makes the occupied-occupied and unoccupied-unoccupied blocks of the Fock matrix diagonal (the off-diagonal blocks are non-zero until convergence).
In such choice of the epoch basis $\mathrm{B}_0$ is defined as the regularized nonzero unique elements on the diagonal of $\mathbf{B}_\mathrm{1e}$:
\begin{align}
(\mathbf{B}_0)_{(ia)(ia)} = 2 n_i r(F_{aa} - F_{ii});
\label{eq:diag_hessian}
\end{align}
the rest of the unique diagonal elements are set to 1 to ensure the finite condition number of the Hessian and existence of the inverse Hessian in an arbitrary basis.
Regularizer $r(x)$ in \cref{eq:diag_hessian} is defined as
\begin{align}
r(x) = \begin{cases}
         x, & \mathrm{if} \quad x > t_r, \\
         t_r & \mathrm{otherwise}.
         \end{cases}
         \label{eq:regularizer}
\end{align}
with the regularizer threshold $t_r$ defining the minimum acceptable HOMO-LUMO gap at the beginning of the epoch.
Unlike some other quasi-Newton solvers, \cite{VRG:vanvoorhis:2002:MP} we do not update the diagonal part of the approximate Hessian every iteration.
In principle, this could lead to slower convergence, since the approximation becomes less accurate as the orbitals are changed from the point where the diagonal Hessian was calculated. \cite{VRG:vanvoorhis:2002:MP}
Indeed, we found that in the early iterations, it is imperative to use an updated preconditioner, and thus we do an approximate line search along the preconditioned steepest descent direction until the max element of the gradient drops below a threshold (we generally use 0.1 which is smaller than 0.25, which has literature precedent \cite{VRG:chaban:1997:TCA}).
During this early phase of the solver, the orbitals are made ``pseudocanonical'' in each iteration, and the preconditioner is rebuilt.
Essentially, the epochs are only 1 iteration long.
However, near the solution, we have found that it is not necessary to update the preconditioner every iteration, and because we work in the epoch MO basis it would be difficult to update the preconditioner.
We have found that with a good initial guess, only a median of 3 iterations of this line search are required to drop the max gradient element below 0.1 and trigger L-BFGS starting for simple systems (see \cref{subsection:g2_data}).
If the gradient gets large again, the history is reset, and preconditioned steepest descent is again carried out with an updated preconditioner.
Every time the history is reset the epoch basis is also reset to the current orbitals.

An alternative and perhaps more conventional view of the preconditioner is that it is a basis transformation that makes the diagonal part of the L-BFGS Hessian or its inverse closer to an identity matrix.
To see how this view relates to the diagonal Hessian, consider the following transformation of the quasi-Newton equation: $\mathbf{s} = -\mathbf{H} \mathbf{g}$ (omitting iteration index).
Multiply both sides of the equation by $\mathbf{B}_{0}^{\mathrm{1/2}}$ and insert $\mathbf{1} = \mathbf{B}_{0}^{\mathrm{1/2}}\mathbf{B}_{0}^{\mathrm{-1/2}}$ (which is clearly acceptable because the diagonal matrix $\mathbf{B}_{0}$ is guaranteed to be positive definite due to the regularizer) between the inverse Hessian and the gradient, to get an equivalent equation.
\begin{align}
\mathbf{B}_{0}^{1/2}\mathbf{s}=-(\mathbf{B}_{0}^{1/2}\mathbf{H} \mathbf{B}_{0}^{1/2})\mathbf{B}_{0}^{-1/2}\mathbf{g}.
\label{eq:qn_step_v0}
\end{align}
With $\tilde{\mathbf{s}}\equiv\mathbf{B}_{0}^{1/2}\mathbf{s}$, $\tilde{\mathbf{g}}\equiv\mathbf{B}_{0}^{-1/2}\mathbf{g}$, and $\tilde{\mathbf{H}}\equiv\mathbf{B}_{0}^{1/2}\mathbf{H}\mathbf{B}_{0}^{1/2}$ identified as the step, gradient, and inverse Hessian, respectively, in the ``preconditioned'' basis (henceforth denoted by the tilde), the Newton step (\cref{eq:qn_step_v0}) becomes
\begin{align}
\tilde{\mathbf{s}} = - \tilde{\mathbf{H}} \tilde{\mathbf{g}}.
\label{eq:newton-step-preconditioned}
\end{align}
The L-BFGS Hessian and inverse Hessian in the preconditioned basis simplify to
\begin{align}
\tilde{\mathbf{B}}_{\mathrm{BFGS}}\equiv \mathbf{B}_{0}^{-1/2}\mathbf{B}_{\mathrm{BFGS}}\mathbf{B}_{0}^{-1/2} = \mathbf{1} + \tilde{\mathbf{V}} (\mathbf{W})^{-1} \tilde{\mathbf{V}}^{\mathrm{T}},
\label{eq:B-BFGS-preconditioned}
\\
\tilde{\mathbf{H}}_{\mathrm{BFGS}}\equiv \mathbf{B}_{0}^{1/2}\mathbf{H}_{\mathrm{BFGS}}\mathbf{B}_{0}^{1/2} = \mathbf{1} + \tilde{\mathbf{V}} \mathbf{M} \tilde{\mathbf{V}}^{\mathrm{T}},
%= \mathbf{B_{0}}^{1/2}(\mathbf{B}_{0}^{-1} + \mathbf{V}_{\mathrm{H}} \mathbf{M} (\mathbf{V}_{\mathrm{H}})^{\mathrm{T}})\mathbf{B_{0}}^{1/2} \nonumber \\
%= & (\mathbf{B}_{0}^{1/2}\mathbf{B}_{0}^{-1}\mathbf{B}_{0}^{1/2}) + \mathbf{B}_{0}^{-1/2}(\mathbf{V}_{\mathrm{H}} \mathbf{M} (\mathbf{V}_{\mathrm{H}})^{\mathrm{T}})\mathbf{B}_{0}^{-1/2} \nonumber \\
%= & \mathbf{1} + \tilde{\mathbf{V}} \mathbf{M} \tilde{\mathbf{V}}^{\mathrm{T}},
\label{eq:H-BFGS-preconditioned}
\end{align}
where
\begin{align}
\tilde{\mathbf{V}}\equiv\mathbf{B}_{0}^{1/2}\mathbf{V}_{\mathrm{H}} = \mathbf{B}_{0}^{-1/2}\mathbf{V}_{\mathrm{B}}.
\end{align}
The $\tilde{\mathbf{V}}^{(k)}$ matrix is computed straightforwardly from
the history vectors in the preconditioned basis.
Note that some steps of the algorithm require quantities in the original basis, such as the sanity checks and computing the orbital rotation matrices via \cref{eq:unitary_step}, thus it is not possible to work exclusively in the preconditioned basis. Transforming back to the original basis is straightforward, e.g., $\mathbf{s}^{(k)}=\mathbf{B}_{0}^{-1/2}\tilde{\mathbf{s}}^{(k)}$.

Here is probably a good place to summarize the steps to obtain the unitary rotation at iteration $k$, since there are now quite a few layers.
\begin{align}
\tilde{\mathbf{s}}^{(k)} \rightarrow \mathbf{s}^{(k)} \rightarrow \bm{\sigma}_{\mathrm{epoch}}^{(k)} \rightarrow \bm{\sigma}^{(k)} \rightarrow \mathbf{U}^{(k)}
\end{align}

To keep the L-BFGS Hessian positive definite between iterations, we require that\cite{VRG:burdakov:2017:MPC}
\begin{align}
\mathbf{s}^{(k)} \cdot \mathbf{y}^{(k)} > t_h ||\mathbf{s}^{(k)}|| \hspace{3pt} ||\mathbf{y}^{(k)}||.
\end{align}
When this requirement is not met vectors $\{\tilde{\mathbf{s}}^{(k)},\tilde{\mathbf{y}}^{(k)}\}$ are not added to the history.
Here we used $t_h = 10^{-5}$.

\subsubsection{Trust-Region Step Restriction.}
\label{subsection:trust_region}

Since the quasi-Newton methods use a quadratic approximation to the objective function, when optimizing a nonlinear function every proposed quasi-Newton step must be tested for sanity to ensure that the quadratic model is a faithful approximation.
First, we expect each step to lower the energy, hence each proposed step should point downhill. Second, steps in downhill directions should not be too large, due to the increasing likelihood that the quadratic model becomes poor.
QUOTR uses the trust-region method for step restriction. In the TR method the maximum step size is limited by the trust-radius that is dynamically updated by comparing the quadratic model predictions with the actual objective function values encountered during the optimization.
Namely step $\tilde{\mathbf{s}}^{(k)}$ is TR-acceptable if
\begin{align}
||\tilde{\mathbf{s}}^{(k)}|| \leq \Delta^{(k)},
\label{eq:trust_region}
\end{align}
where trust-radius $\Delta^{(k)}$ is updated using
Fletcher's algorithm.\cite{VRG:fletcher:1980:}
By comparing the quadratic model prediction for the energy with the actual energy value every iteration the trust-radius can be expanded, contracted, or left unchanged (see \cref{algorithm:scf_solver}).
Fletcher's algorithm parameters ($\tau_i$/$\eta_i$) were borrowed from a recent study. \cite{VRG:burdakov:2017:MPC}
The initial value of the trust-radius is set to the most recent successful line search step size since this should be of the correct order of magnitude for the next step. 
Also, we always perform line search when there are no history data available so the most recent step size from line search is always a known quantity when quasi-Newton steps are attempted.

When the quasi-Newton step does not satisfy \cref{eq:trust_region}, we solve for the optimal step that is within the TR, which must be on the TR boundary: $||\tilde{\mathbf{s}}^{(k)}|| = \Delta^{(k)}$. 
This is done by finding an optimal level-shift, $\sigma$, which satisfies the two conditions:\cite{VRG:burdakov:2017:MPC}
\begin{subequations}
\begin{align}
(\tilde{\mathbf{B}}_{\mathrm{BFGS}} + \sigma \mathbf{1}) \tilde{\mathbf{s}}^{(k)} = & -\tilde{\mathbf{g}}^{(k)}
\label{eq:tr_condition1}
\\
    ||\tilde{\mathbf{s}}^{(k)}|| = \Delta^{(k)}.
    \label{eq:tr_condition2}
\end{align}
    \label{eq:tr_conditions}
\end{subequations}
Level-shift $\sigma$ of the L-BFGS Hessian is updated iteratively until \cref{eq:tr_conditions} is satisfied to the desired precision controlled by parameters $T_{1,2}$ (see \cref{table:tr_parameters,algorithm:tr_solver}).

The TR solver in QUOTR is based on the solver described by Burdakov {\em et al.}\cite{VRG:burdakov:2017:MPC} that leverages the low-rank structure of the L-BFGS Hessian. The advantage of this approach is that the cost of each TR solver iteration is trivial compared to the conventional TR formulation with exact Hessian in which each iteration has a cost similar to that of the gradient evaluation.
The TR solver algorithm is outlined in \cref{algorithm:tr_solver} and its user-controllable parameters are given in \cref{table:tr_parameters}.
Although the parameters for the TR solver could be adjusted, we recommend keeping them as listed.
The initial guess for $\sigma$ is best kept at zero, as this will help convergence in cases where the optimal level shift is a very small value.
Changing parameters $T_{1,2}$ only impacts when the TR is problem is considered solved, and tightening the values given is not expected to have a significant impact on QUOTR overall.

\begin{algorithm}
\caption{Trust-Region Step Update}
\begin{algorithmic}[1]

\Function{TRStep}{$\Delta,\tilde{\mathbf{g}},\tilde{\mathbf{s}}, \mathbf{\tilde{V}}$}

\State $\sigma \gets \sigma_\mathrm{init}$, $C \gets \code{false}$, $F \gets F_0$ \Comment{initialize}

\State $\mathbf{G} \gets \tilde{\mathbf{V}}^{\mathrm{T}}\tilde{\mathbf{V}}$
\State \{$\mathbf{R}, \mathbf{R}^{-1}\} \gets \Call{Orthogonalize}{\mathbf{G}}$ \Comment{\cref{eq:orthogonalizer}}
\State $\mathbf{R}^{-1} \mathbf{W}^{-1} (\mathbf{R}^{-1})^{\mathrm{T}} = \mathbf{U} \Lambda \mathbf{U}^{\mathrm{T}}$ \Comment{diagonalize}
\State $\mathbf{P}_{\parallel} = \tilde{\mathbf{V}} \mathbf{R} \mathbf{U}$ \Comment{Eq. 12\cite{VRG:burdakov:2017:MPC}}
\State $\tilde{\mathbf{g}}_{\parallel} = (\mathbf{P}_{\parallel})^{\mathrm{T}} \tilde{\mathbf{g}}$
\State $r_{\mathrm{max}} = \mathrm{dim}(\tilde{\mathbf{g}}_{\parallel})$
\State $||\tilde{\mathbf{g}}_{\perp}||^{2} = ||\tilde{\mathbf{g}}||^{2} - ||\tilde{\mathbf{g}}_{\parallel}||^{2}$ \Comment{Eq. 25\cite{VRG:burdakov:2017:MPC}}
\State $i = 0$

\While{not $C$ AND $i < i_{\mathrm{max}}$}
\State $\tilde{\mathbf{v}}_{\parallel} \gets -(\Lambda + \sigma \mathbf{1})^{-1} \tilde{\mathbf{g}}_{\parallel} $ \Comment{Eq. 16\cite{VRG:burdakov:2017:MPC}}
\State $||\tilde{\mathbf{v}}|| \gets (||\tilde{\mathbf{v}}_{\parallel}||^{2} + ||\tilde{\mathbf{g}}_{\perp}||^{2}/(1 + \sigma)^{2})^{1/2}$ \Comment{Eq. 20\cite{VRG:burdakov:2017:MPC}}
\State $v_{\mathrm{temp}} \gets -||\tilde{\mathbf{g}}_{\perp}||^{2}/(1 + \sigma)^{3}$ \Comment{Eq. 21\cite{VRG:burdakov:2017:MPC}}
\State $r \gets 0$
\While{$r < r_{\mathrm{max}}$}
\State $v_{\mathrm{temp}} \gets v_{\mathrm{temp}} - (\tilde{\mathbf{g}}_{\parallel}[r])^{2}/(1 + \sigma)^{3}$ \Comment{Eq. 21\cite{VRG:burdakov:2017:MPC}}
\State $r \gets r + 1$
\EndWhile
\State $\sigma \gets \sigma - \frac{\phi(\sigma)}{\phi'(\sigma)} = \sigma - \frac{(||\tilde{\mathbf{v}}|| - \Delta)||\tilde{\mathbf{v}}||^{2}}{v_{\mathrm{temp}}\Delta}$ \Comment{Eq. 19\cite{VRG:burdakov:2017:MPC}}

\If{$\sigma \in \{\sigma_{-1}, \sigma_{-2}, \sigma_{-3}, \sigma_{-4}\}$} \Comment{stabilize}
\State $\sigma \gets \frac{1}{2}(\sigma + \sigma_{-1})$
\EndIf
\State $\{\sigma_{-1}, \sigma_{-2}, \sigma_{-3}, \sigma_{-4}\} \gets \{\sigma, \sigma_{-1}, \sigma_{-2}, \sigma_{-3}\}$

\If{$|||\tilde{\mathbf{v}}|| - \Delta| \leq \mathrm{min}(T_1 \Delta, T_2)$} \Comment{converged?}
\If{$\sigma < 0$} \Comment{wrong sign?}
\State $\sigma \gets -\sigma F$ \Comment{reset $\sigma$}
\State $F \gets F F_0$
\State $C =$ \code{false}
\ElsIf{$\sigma \geq 0$}
\State $C =$ \code{true}  \Comment{converged}
\EndIf

\EndIf

\State $i \gets i + 1$

\EndWhile

\State $\tilde{\mathbf{v}}_{\parallel} \gets -(\Lambda + \sigma \mathbf{1})^{-1} \tilde{\mathbf{g}}_{\parallel} $ \Comment{Eq. 16\cite{VRG:burdakov:2017:MPC}}
\State $\tilde{\mathbf{s}} = \mathbf{P}_{\parallel} (\tilde{\mathbf{v}}_{\parallel} + (1 + \sigma)^{-1}\tilde{\mathbf{g}}_{\parallel}) - (1 + \sigma)^{-1}\tilde{\mathbf{g}}$ \Comment{Eq. 27\cite{VRG:burdakov:2017:MPC}}
\If{$i \geq i_{\mathrm{max}}$}
\State $\tilde{\mathbf{s}} \gets \tilde{\mathbf{s}}_{\mathrm{given}}$ \Comment{use original step}
\EndIf

\State \Return $\tilde{\mathbf{s}}$
\EndFunction

\end{algorithmic}
\label{algorithm:tr_solver}
\end{algorithm}

\begin{table}[h]
    \caption{User-controllable parameters of the TR solver.}
    \label{table:tr_parameters}
    \begin{tabular*}{0.48\textwidth}{@{\extracolsep{\fill}}lll}
        \hline
        Description & Symbol & Value \\
        \hline
        Initial guess $\sigma$ & $\sigma_\mathrm{init}$ & 0.0 \\
        Convergence criterion 1 & $T_1$ & $10^{-4}$ \\
        Convergence criterion 2 & $T_2$ & $10^{-7}$ \\
        maximum iterations & $i_{\mathrm{max}}$ & 500 \\
        sigma modify factor & $F_0$ & 1.1 \\ 
        \hline
    \end{tabular*}
\end{table}

We have modified the algorithm of Ref. \citenum{VRG:burdakov:2017:MPC} in several ways.
First, the use of rank-revealing Cholesky decomposition in Ref. \citenum{VRG:burdakov:2017:MPC} (see text around their Eq. (9)) is replaced by the use of L\"owdin canonical orthogonalization.\cite{VRG:lowdin:1956:AiP}
Inserting \cref{eq:B-BFGS-preconditioned} in \cref{eq:tr_condition1} produces
\begin{align}
((1+\sigma) \mathbf{1} + \tilde{\mathbf{V}} (\mathbf{W})^{-1} \tilde{\mathbf{V}}^{\mathrm{T}}) \tilde{\mathbf{s}}^{(k)} = & -\tilde{\mathbf{g}}^{(k)}.
\end{align}
Burdakov {\em et al.} use rank-revealing Cholesky decomposition of the history Gramian $\mathbf{G} \equiv \tilde{\mathbf{V}}^\mathrm{T} \tilde{\mathbf{V}}$. Here we obtain matrix $\mathbf{R}$ satisfying
\begin{align}
\mathbf{R}^\mathrm{T} \mathbf{G} \mathbf{R} = & \mathbf{1}
\end{align}
by canonical orthogonalization ignoring Gramian eigenvalues less than $\epsilon$:
\begin{subequations}
\begin{align}
    \mathbf{G} = & \mathbf{U} \mathbf{g} \mathbf{U}^T, \quad g_i >\epsilon \label{eq:diag_gramian}\\
    \mathbf{R} = & \mathbf{U} \mathbf{g}^{-1/2} \label{eq:define_r}\\
    \mathbf{R}^{-1} = & (\mathbf{U} \mathbf{g}^{1/2})^{\mathrm{T}}
    \label{eq:define_r_inv}
\end{align}
\label{eq:orthogonalizer}
\end{subequations}
This allows us to implement the algorithm more portably, using only standard linear algebra available in LAPACK. Other differences can be seen in the algorithm listing, but the most notable changes include:
\begin{itemize}
\item we enforce $\sigma > 0$,
\item we prevent some infinite loops by averaging $\sigma$ with the previous value if that value of $\sigma$ occurred in the last four iterations,
\item we simplify convergence criteria to merely check that step size is within a threshold difference of $\Delta$.
\end{itemize}

Once a TR-compliant step has been determined 
energy (and gradient) is evaluated at the displaced geometry and
compared to the quadratic model estimate \begin{align}
q^{(k)} = \tilde{\mathbf{s}}^{(k)} \cdot \tilde{\mathbf{g}}^{(k)} + \frac{1}{2}\tilde{\mathbf{s}}^{(k)} \cdot \tilde{\mathbf{B}}_{\mathrm{BFGS}}^{(k)} \tilde{\mathbf{s}}^{(k)}.
\label{eq:qk}
\end{align}
If the actual energy change differs too much from $q^{(k)}$ (based on $\tau_1$ and $t_0$), the step is rejected, the trust-radius is decreased and used to update the step by re-solving the TR problem.
If the quadratic model is catastrophically bad, repeated shrinking of TR may occur.
The lower limit for TR, $t_t$, plays the role of an escape hatch for such a scenario; if TR becomes smaller than $t_t$, we reset the history, do a single line search iteration, and continue from there with the new trust-radius determined from the line search step size.

Note that the TR problem is always solved in the preconditioned epoch basis; this is yet another reason to use the same basis throughout the epoch. When a new epoch starts and the preconditioner and the epoch basis change we cannot simply carry over the TR value between epochs.
Thus, however, since each epoch starts with a line search step, this produces a fresh initial estimate of the trust-radius valid in that epoch's preconditioned basis.

\subsubsection{Line Search}
\label{sec:line-search}

Each QUOTR epoch starts with a single steepest descent step:
\begin{align}
\mathbf{\tilde{s}}^{(k)} \overset{\mathrm{SD}}{=} - \alpha^{(k)} \mathbf{\tilde{g}}^{(k)} / ||\mathbf{\tilde{g}}^{(k)}||.
\end{align}
Step ``size'' $\alpha^{(k)}$ is determined by a line search.  To reduce the number of gradient evaluations we use an {\em approximate} line search. First, the energy along the SD direction $E(\alpha)$ is approximated by its 3rd-order polynomial $E_\mathrm{fit}(\alpha)$ on a fitting interval $[0,\alpha_{\mathrm{fit}})$:
\begin{align}
E(\alpha) \approx E_\mathrm{fit}(\alpha) = a \alpha^3 + b \alpha^2 + c\alpha + d.
\end{align}
Coefficients $\{a,b,c,d\}$ are determined by matching exactly the energies and gradients evaluated with the current orbitals ($\alpha=0$) and at the end of the fitting interval $\alpha_{\mathrm{fit}}$ (see below). Thus at the beginning of QUOTR 2 gradient evaluations are required for the line search; in subsequent epochs only 1 extra gradient evaluation is needed since the current orbitals' gradient has been computed as part of the previous epoch.
A similar procedure has recently been used, \cite{VRG:ivanov:2021:CPC} and a nearly identical method for the step determination also has precedent. \cite{VRG:stanton:1981:JCP}

Choosing $\alpha_\mathrm{fit}$ is crucial for the success of the line search. Due to the fact that the exponential parametrization of the unitary is periodic, it is straightforward to estimate the shortest period of oscillation by finding the largest (magnitude) eigenvalue $\omega_{\mathrm{max}}$ of the $\rm{\sigma}$ matrix (or matrices, for the unrestricted case) corresponding to the SD direction $-\mathbf{\tilde{g}}^{(k)}/||\mathbf{\tilde{g}}^{(k)}||$.\cite{VRG:abrudan:2009:SP}
This results in
\begin{align}
\alpha_{\mathrm{fit}} = \frac{2\pi}{q|\omega_{\mathrm{max}}|}
\end{align}
Here $q$ is set to 4 due to the quartic dependence of the Hartree-Fock energy on the orbitals without the orthonormality constraint.

The minimum of $E_\mathrm{fit}(\alpha)$ is found by solving the quadratic equation $d E_\mathrm{fit}/d \alpha = 0$. Each of its two solutions, $\alpha_\mathrm{min}$, is checked for sanity in turn; it is expected to be real, positive, resulting in a decrease of energy (i.e., $E_\mathrm{fit}(\alpha_\mathrm{min}) < E_\mathrm{fit}(0)$), and the second derivative of $E_\mathrm{fit}$ at $\alpha_\mathrm{min}$ should be positive.
Since a 3rd-order polynomial can have at most one local minimum, the solution with the smallest positive $\alpha$ is checked whether it satisfies all of these criteria. The failure to meet these conditions indicates a poor quality of the fit, and in such case the polynomial fit is recomputed with $\alpha_\mathrm{fit}$ scaled by $\alpha_\mathrm{fit,shrink}=1/2$.
Note that the energy decrease criterion is checked only after building the new Fock matrix, while the other conditions can be checked immediately after the solving the quadratic equation.
    
\subsubsection{Orbital guess.}

Starting (guess) orbitals are another critical component for rapid SCF convergence.
We generally use an extended H\"{u}ckel initial guess,\cite{VRG:hoffmann:1963:JCP} which is constructed in a minimal basis and then projected onto the full orbital basis. 
The standard Wolfsberg-Helmholtz formula for the off-diagonal elements of the extended H\"{u}ckel Hamiltonian is used:\cite{VRG:hoffmann:1963:JCP} $H_{ij} = K' S_{ij} (H_{ii} + H_{jj})/2$, but with the updated formula for the value of $K'$.\cite{VRG:ammeter:1978:JACS} Instead of experimental ionization potentials for the diagonal elements, we follow a suggestion by Lehtola \cite{VRG:lehtola:2019:JCTC} (earlier by Norman\cite{VRG:norman:2012:CPL}) and use numerical Hartree-Fock orbital energies\cite{VRG:fischer:1972:ADaNDT} for each shell.
Although the guess orbitals are populated according to the Aufbau principle using the extended H\"{u}ckel energies and in the minimal basis, after projection to the orbital basis the populations may not be qualitatively correct.
When this situation occurs, and the symmetry of the orbitals is such that there is no gradient between incorrectly occupied and incorrectly unoccupied orbitals, QUOTR will not be able to correct the populations.
Therefore, we have added an option to perturb the guess orbitals to allow the solver to rotate the incorrectly occupied orbitals and find the lower energy solution.
The orbitals are perturbed by $\exp(\bm{\sigma})$ with unique elements of $\bm{\sigma}$ filled with uniformly-distributed random numbers in $[-0.05,0.05]$.
The ``strength'' of this random perturbation can be changed, which simply scales all elements of $\bm{\sigma}$ such that the maximum absolute value is something other than 0.05 (default).
Additionally, we can choose to either perturb ``All'' or just the ``Valence'' orbitals.
Pseudorandom number generator is used with user-controlled seed to ensure deterministic perturbation.

\subsubsection{Additional Heuristics.}

Unfortunately, the sole use of preconditioned L-BFGS with TR step restriction is not sufficient when dealing with problems with complex optimization landscapes that arise for open-shell and, especially, metal-containing systems. This occurs due to poor quality of the quadratic model when far from convergence. Thus, as is typical with second-order solvers,\cite{VRG:chaban:1997:TCA,VRG:vanvoorhis:2002:MP} QUOTR uses SD steps until the $||\mathbf{g}||_\infty$ drops below the L-BFGS start threshold, $t_b$.

Here's a brief recap of all situations that cause history reset:
\begin{itemize}
    \item The gradient is too large ($||\mathbf{g}^{(k)}||_{\infty} > t_b$)
    \item The TR is too small ($\Delta^{(k)} < t_t$)
    \item The quadratic model predicts energy increase ($q^{(k)} > 0$)
\end{itemize}
The first 2 of these situations are detected before the L-BFGS step is constructed, allowing the solver to skip this step to go directly to re-building the preconditioner and on to perform the line search.
The last situation is only determined after the L-BFGS step is calculated.

\section{Technical Details}
\label{section:details}

The QUOTR solver was implemented in a developmental version of the Massively Parallel Quantum Chemistry (MPQC) version 4 program package.\cite{VRG:peng:2020:JCP}
The default values for parameters in \cref{table:scf_parameters,table:tr_parameters} were used throughout, unless noted.
The orbital bases sets used were 6-31G*,\cite{VRG:ditchfield:1971:JCP,VRG:hehre:1972:JCP,VRG:hariharan:1973:TCA,VRG:dill:1975:JCP,VRG:binkley:1977:JCP,VRG:gordon:1982:JACS,VRG:francl:1982:JCP} 6-31G**,\cite{VRG:hariharan:1973:TCA} 6-311++G**,\cite{VRG:krishnan:1980:JCP,VRG:mclean:1980:JCP,VRG:clark:1983:JCC,VRG:spitznagel:1987:JCC} def2-TZVPP,\cite{VRG:weigend:2005:PCCP} cc-pVTZ-DK,\cite{VRG:dejong:2001:JCP} and cc-pVTZ-X2C.\cite{VRG:feng:2017:JCP}
Density fitting, where noted, used the def2-universal-J basis. \cite{VRG:weigend:2006:PCCP}
The extended H\"{u}ckel initial guess was constructed in the Huzinaga MINI basis, \cite{VRG:schuchardt:2007:JCIM} then projected onto the orbital basis.
Calculations on the f-element containing system in \cref{subsection:difficult_f_systems} did not use the extended H\"uckel guess to avoid the uncertainties about its quality in such heavy systems. Instead, we use a superposition of minimal atomic basis guess densities to construct the initial Fock matrix in the orbital basis (without projection), followed by diagonalization. The minimal AO basis used the corresponding subset of the ANO-DK3 basis\cite{VRG:tsuchiya:2001:JCP} on Fm atom and the MINI AO basis on the other atoms.
The same minimal bases were used to compute the atomic charges in \cref{subsection:difficult_f_systems}, using the pseudoinverse method described in Ref. \citenum{VRG:clement:2021:JCTC}. 
The orbital bases used in the relativistic calculations employed cc-pVTZ-X2C on the Fm atom, and cc-pVTZ-DK on all other atoms.

Hartree-Fock calculations were performed in \cref{subsection:g2_data} for the G2 set, \cite{VRG:curtiss:1997:JCP} the geometries for which were obtained from the Gaussian output files on the NIST website \cite{VRG:johnson:2002:} with the exception of four systems that were not available with the correct method (MP2=FULL/6-31G*).
For the four systems that were not available from NIST (acetamide, furan, SiH$_2$-triplet and 2-butyne) Gaussian 09\cite{VRG:frisch:2013:} was used to obtain the geometry.
The G2-1 set consists of 55 systems and is a subset of G2-2, which consists of a total of 148 systems. \cite{VRG:curtiss:1997:JCP}

Henceforth RH/DIIS will be denoted simply by DIIS. Unless explicitly mentioned, DIIS results were obtained with its implementation in MPQC using the default parameters: keeping the 5 most recent pairs of Fock matrix and error vectors for the extrapolation, and no damping applied.

The KS DFT implementation in MPQC uses GauXC\cite{VRG:petrone:2018:EPJB} (which uses LibXC\cite{VRG:lehtola:2018:S}) for calculation of the exchange-correlation potentials and energies.
The integration grid used for the 1PLW calculations in \cref{subsection:vanishing_gap_system} was the ``ultrafine'' grid (99 radial Mura-Knowles \cite{VRG:mura:1996:JCP} points, 590 angular Lebedev-Laikov \cite{VRG:lebedev:1999:DM} points). 
All other KS DFT calculations used the ``superfine'' grid which has 250 radial points and 974 angular points for all atoms except hydrogen, which has 175 radial points.
The particular parameterization we use for LDA is Slater Exchange\cite{VRG:dirac:1930:MPCPS} with VWN RPA. \cite{VRG:vosko:1980:CJP}
For the B3LYP calculations on the Cr systems in \cref{subsection:difficult_d_systems}, we use VWN3 for the local correlation functional\cite{VRG:vosko:1980:CJP} to match PySCF.
The structure of the neuropeptide, 1PLW, \cite{VRG:marcotte:2004:BJ} was obtained from the Protein Data Bank (PDB). \cite{VRG:berman:2000:NAR}

Calculations using KDIIS\cite{VRG:kollmar:1997:IJQC} for SCF acceleration on the CrC and Cr$_2$ systems in \cref{subsection:difficult_d_systems} were performed with the Orca program system, version 5.0.4.\cite{VRG:neese:2022:WCMS}
Additionally, the DIIS implementation from Orca was also used for these systems instead of the MPQC version.
The bond length used for both of these diatomic systems is 2 angstrom, as has been used in previous studies.\cite{VRG:daniels:2000:PCCP,VRG:helmich-paris:2021:JCP}
Orbitals were plotted for Cr$_2$ with Jmol; due to its inability to read in Molden files with $l=4$ (g) AOs, the calculations (only for the visualizations) were performed with def2-TZVPP with g-type AOs removed.

Full (2-component) and spin-free (1-component) 1-electron X2C Hamiltonians were implemented in MPQC using the standard formalism.\cite{VRG:liu:2009:JCP,VRG:peng:2013:JCP}
No empirical scaling was utilized to emulate the mean-field effects on the Dirac Hamiltonian.
For the sake of comparison with the results of Ref. \citenum{VRG:penchoff:2018:AO} only the spin-free X2C Hamiltonian was used here.

\section{Results and Discussion}
\label{section:results}

\subsection{Easy Testset: G2 Data Set}
\label{subsection:g2_data}

Performance of QUOTR was first assessed for converging Hartree-Fock wave functions (RHF and UHF for closed- and open-shell systems, respectively) and compared to DIIS, as well as published literature data for three second-order solvers, GDM,\cite{VRG:vanvoorhis:2002:MP}
ETDM,\cite{VRG:ivanov:2021:CPC} and
CIAH.\cite{VRG:sun:2017:AP}
The computations where convergence was not achieved in 256 iterations (333 for ETDM, 50 macroiterations for CIAH) were removed from the statistical values and aggregated in  the ``no convergence'' row.
The number of ``local minima'' for each solver was determined by comparing converged energies to the lowest energy that we obtained, except for the GDM and ETDM results (which are the numbers reported by the respective publications).
All calculations with QUOTR and DIIS used the extended H\"{u}ckel guess with perturbed valence orbitals.
Although we attempted to compare solvers as faithfully as possible, due to lack of direct access to the source code and/or implementation of GDM and ETDM this was not always possible (see below).

\cref{table:g2_stats_compare} reports the number of Fock matrix evaluations $N_\mathrm{F}$ (``Fock builds'') and the number of solver iterations $N_\mathrm{I}$. Due to the different performance statistics reported in the literature for GDM and ETDM, three different sets of G2 calculations were performed.
\begin{table*}
    \caption{Performance comparison of QUOTR to other SCF solvers for standard G2 set.}
    \begin{tabular*}{\textwidth}{@{\extracolsep{\fill}}lccccccccc}
    \hline\hline
    {} & \multicolumn{3}{c}{G2-1/6-311++G**} & \multicolumn{3}{c}{G2-2/6-31G**} & \multicolumn{3}{c}{G2-2/6-31G*} \\ \cmidrule(lr){2-4} \cmidrule(lr){5-7} \cmidrule(lr){8-10}
    {} & DIIS & QUOTR & GDM$^a$ & DIIS & QUOTR & ETDM$^b$ & DIIS & QUOTR & CIAH$^c$ \\
    \hline

    $N_\mathrm{F}$: mean & 15.2 & 26.8 & \textemdash & 13.6 & 20.5 & (17) & 15.3 & 19.4 & 34.5 \\
    $N_\mathrm{F}$: median & 14 & 20 & \textemdash & 12 & 17 & (17) & 12 & 16 & 30 \\
    $N_\mathrm{F}$: max & 40 & 136 & \textemdash & 64 & 107 & 72 & 234 & 69 & 77 \\
    $N_\mathrm{I}$: mean & 15.2 & 20.5 & 16.3 & 13.6 & 15.3 & \textemdash & 15.3 & 14.2 & 2.9 \\
    $N_\mathrm{I}$: median & 14 & 15 & \textemdash & 12 & 12 & \textemdash & 12 & 12 & 3 \\
    $N_\mathrm{I}$: max & 40 & 107 & 42 & 64 & 96 & \textemdash & 234 & 55 & 5 \\
    \hline
    local minima & 3 & 0 & 5 & 5 & 0 & \textemdash & 8 & 0 & 4 \\
    no convergence & 2 & 0 & 0 & 2 & 0 & 0 & 1 & 0 & 1 \\    
    \hline\hline
    \end{tabular*}
    
    \begin{tablenotes}
    \item $^a$ Geometric Direct Minimization \cite{VRG:vanvoorhis:2002:MP}
    \item $^b$ Exponential Transformation Direct Minimization \cite{VRG:ivanov:2021:CPC}
    \item $^c$ Co-iterative augmented Hessian \cite{VRG:sun:2017:AP}
    \end{tablenotes}
    
    \label{table:g2_stats_compare}
\end{table*}

\subsubsection{G2-1/6-311++G**.}

Similar to QUOTR, the Geometric Direct Minimization (GDM) solver is a BFGS-based solver introduced by Head-Gordon and Van Voorhis in 2002.\cite{VRG:vanvoorhis:2002:MP}
A key difference between GDM and QUOTR is the use of the TR by QUOTR.
Additionally, GDM updates the preconditioner every iteration (rather than once per epoch in QUOTR), with regularization applied by adding a diagonal shift to the Hessian equal to the energy change in the most recent iteration.
Therefore, comparison to GDM is appropriate as a way to evaluate the effectiveness of the TR and the appropriateness of the preconditioner. Due to the lack of access to the commercial implementation of GDM, we restricted our comparison to the data published in Ref.~\citenum{VRG:vanvoorhis:2002:MP}.

To make the comparison with GDM as faithful as possible, we used the same orbital basis set and convergence criteria ($1 \times 10^{-10} E_{\mathrm{h}}$ for the energy difference between iterations and $1 \times 10^{-7}$ for the RMS of the unique gradient elements).
Our initial guess orbitals were likely similar; however, we did apply a random unitary perturbation to the valence orbitals, which was not done by GDM.
The average number of iterations taken by QUOTR is about 4 more than GDM, and it has a higher max at 107 for NO followed by P$_2$ at 66 iterations.
Notice, though, that GDM found 5 local minima relative to the lowest energy that they could obtain in any of their calculations.
Thus, while the convergence with QUOTR takes more iterations, QUOTR appears to be more robust than GDM.
Unfortunately, the number of Fock builds used by GDM was not reported in Ref. \citenum{VRG:vanvoorhis:2002:MP}.
The computational cost of QUOTR is fairly competitive with DIIS, with a median number of Fock builds being 20 and 14, respectively; note that the latter is close to the reported performance of DIIS in the GDM paper.\cite{VRG:vanvoorhis:2002:MP}
The only systems that did not converge for DIIS in 256 iterations were HCO and Si$_2$.
There were no local minima found by QUOTR (relative to the DIIS solution) but for 3 systems (CN, O$_2$ and CH singlet) QUOTR got a significantly lower energy than DIIS.
Also note that for 5 systems GDM landed on local minima too.\cite{VRG:vanvoorhis:2002:MP}
Thus for the G2-1/6-311++G** test set QUOTR was found to be more robust than DIIS and GDM, albeit with a slightly worse performance.

\subsubsection{G2-2/6-31G**.}

Another similar, and more recent, direct minimization SCF solver is the exponential transformation direct minimization (ETDM) solver.\cite{VRG:ivanov:2021:CPC}
Again, the lack of TR is one of the main differences compared to QUOTR, but ETDM also approximates the gradient as it does not work in the epoch formulation.
The fact that ETDM does not use TR may be compensated by the stronger criteria used in the line search.  Due to the lack of access to the implementation of ETDM we restricted our comparison to the data published in Ref.~\citenum{VRG:ivanov:2021:CPC}.

Comparison to ETDM was less precise for a few reasons.
The data presented in \cref{table:g2_stats_compare} for ETDM used KS DFT (PBE) and a different basis (double-zeta polarized numerical atomic orbital basis equipped with projector augmented wave (PAW) for the inner region).
Here we performed all-electron SCF in the 6-31G** Gaussian AO basis, because it is a double-zeta basis with polarization functions on all atoms, so should be similar to the basis used by ETDM. Also, the open-shell KS DFT implementation in MPQC is not yet finalized, hence we are comparing QUOTR HF SCF to ETDM PBE SCF.
Lastly, QUOTR set $m = 3$ to make the comparison with ETDM as faithful as possible.
The average number of iterations (unclear if it is mean or median) reported for ETDM was 17, which compares well with the median of QUOTR at 17.
Therefore, we conclude that QUOTR is roughly equivalent in computational cost to ETDM.
Notice also that 5 of the DIIS solutions are local minima relative to QUOTR.

\subsubsection{G2-2/6-31G*.}

The last batch of comparisons pits QUOTR and DIIS against CIAH, a second-order (augmented Hessian) solver that uses exact Hessian.\cite{VRG:sun:2017:AP} 
The augmented Hessian (AH) approach can be viewed as a variant of the Newton method with step restriction induced by  a spectral shift of the Hessian tuned at each step to ensure that the predicted step results in energy decrease (this idea is sufficiently general to be applicable in combination with RH/DIIS\cite{VRG:host:2008:JCP}). The spectral shift can be viewed as an  optimal regularizer for the Hessian; since it vanishes automatically in the vicinity of the minimum the augmented Hessian approach approaches the quadratic convergence rate of the unmodified Newton method. Thus the augmented Hessian methods are potentially superior to RH/DIIS or quasi-Newton methods (like QUOTR) that have slower convergence rates. Indeed, for the SCF problem the AH methods are known\cite{VRG:sun:2017:AP,VRG:helmich-paris:2021:JCP} to converge in substantially fewer iterations than the RH/DIIS heuristics, and are more robust.
Thus the AH-based SCF methods can viewed as the benchmark to beat for QUOTR.

With access to the implementation of CIAH in PySCF\cite{VRG:sun:2020:JCP} we were able to perform a direct comparison with QUOTR (\cref{table:g2_stats_compare}). Note that the number of iterations is sometimes used\cite{VRG:dittmer:2023:JCP} to compare the cost of the AH-based methods to that of RH/DIIS and quasi-Newton approaches; such comparison is misleading. Each iteration of the AH approach, in addition to the evaluation of the gradient, involves iteratively solving an eigenproblem defining the optimal shift; thus the cost of each iteration is determined by the cost of multiplying a trial step vector by the (exact) Hessian times the number of iterations of the eigensolver. The cost of applying exact Hessian to the trial step is comparable to the cost of the Fock matrix evaluation (in fact many programs will use the same machinery for both). Thus the performance assessment of CIAH and other AH-based methods will report the number of {\em Fock build equivalents}, $N_\mathrm{F}$.
For CIAH the total number of Fock build equivalents is the sum of the key frames (KF) and coulomb/exchange (JK) calls,
with the former accounting for the cost of the {\em exact} evaluation of the gradient and the latter the cost of Hessian-step products.\cite{VRG:sun:2017:AP}

The convergence statistics in \cref{table:g2_stats_compare} indicate that the median $N_\mathrm{F}$ for QUOTR is roughly half of the median $N_\mathrm{F}$ for CIAH.
QUOTR was also more robust: for one system (HCl) CIAH did not converge as it could not reduce the gradient norm below $1.2 \times 10^{-6}$ and failed to make progress until the maximum number of iterations was reached (50).
In four other systems (CH, $\mathrm{O}_2$, $\mathrm{NO}_2$ and $\mathrm{Si}_2$) CIAH landed on a local minimum, as indicated by the substantially lower energies obtained with QUOTR. For the rest of the systems, CIAH and QUOTR agreed within $1 \times 10^{-9} E_{\mathrm{h}}$.
The systems that took the most Fock build equivalents to converge with QUOTR and CIAH were Si$_2$ and NF$_3$, respectively, requiring 69 and 77.

As expected, RH/DIIS is on average slightly faster than QUOTR, but is far less robust, with 8 local minima and 1 system where converged solution could not be obtained.
This demonstrated proliferation of incorrect solutions found with DIIS for even such ``easy" chemical systems at those in the G2 test set has significant implications.
How could high-throughput screening be performed with confidence using such a solver?
We expect that other programs that default to using the RH solver will also have similar issues.
And, as demonstrated in Ref. \citenum{VRG:vanvoorhis:2002:MP}, second-order solvers are not a panacea.

Even with direct access to the CIAH implementation it was difficult to compare methods fairly. Some differences were minor, such as the convergence tests.
Both solvers use magnitudes of energy change and gradient for convergence monitoring. For the former threshold was set to $1 \times 10^{-9} E_{\mathrm{h}}$, but the CIAH gradient criterion ($1 \times 10^{-6}$) is defined in terms of the gradient norm rather than the RMS value used by QUOTR ($1 \times 10^{-5}$).
Some differences were more significant, like the choice of the guess orbitals. The QUOTR data in  \cref{table:g2_stats_compare} was obtained with its default guess (perturbed extended H\"{u}ckel) that differs from the default minimal AO guess used for CIAH calculations. To elucidate the impact of the guess differences we performed additional tests with (unperturbed) core Hamiltonian guess implemented identically in MPQC and PySCF to ensure that the initial energies matched to better than 9 digits between the two programs.
The convergence criteria for QUOTR were changed to match the criteria of CIAH, using $1 \times 10^{-6}$ norm of the unique elements of the gradient, and the criterion on energy change was kept at $1 \times 10^{-9} E_{\mathrm{h}}$.
The results for RHF computations of 10 small molecules in the G2 set are displayed in \cref{table:g2-10_nFock_compare}. In all cases, CIAH and QUOTR found the same solution but the former required on average 3 times more Fock build equivalents.

\begin{table}
    \caption{Performance comparison of CIAH and QUOTR for a subset of G2-2/6-31G* with core Hamiltonian guess.}
    \begin{tabular*}{0.48\textwidth}{@{\extracolsep{\fill}}ccccc}
    \hline\hline
    {} & \multicolumn{3}{c}{CIAH} & QUOTR \\ \cmidrule(lr){2-4} \cmidrule(lr){5-5}
     System & KF & JK & $N_\mathrm{F}$ & $N_\mathrm{F}$  \\
     \hline
    $\mathrm{CH_4}$ & 9 & 35 & 44 & 14 \\
    $\mathrm{CO}$ & 17 & 70 & 87 &  22 \\
    $\mathrm{F_2}$ & 7 & 24 & 31 &  12 \\
    $\mathrm{H_2}$ & 3 & 7 & 10 & 5 \\
    $\mathrm{H_2 O}$ & 11 & 43 & 54 & 14 \\
    $\mathrm{HF}$ & 12 & 48 & 60 & 13 \\
    $\mathrm{Li_2}$ & 6 & 18 & 24 & 10 \\
    $\mathrm{LiH}$ & 5 & 15 & 20 & 10 \\
    $\mathrm{N_2}$ & 9 & 34 & 43 & 13 \\
    $\mathrm{NH_3}$ & 12 & 53 & 65 & 19 \\
    \hline
    median & 9 & 34.5 & 43.5 & 13 \\
    mean & 9.1 & 34.7 & 43.8 & 13.2 \\
    max & 17 & 70 & 87 & 22 \\ \hline\hline
    \end{tabular*}
    \label{table:g2-10_nFock_compare}
\end{table}

A deeper breakdown of QUOTR's convergence statistics for the G2-2/6-31G* set is presented in \cref{table:stats_overall}, where ``Before L-BFGS'' refers to the line search iterations {\em before} the first use of a quasi-Newton step, and ``Line Search'' refers to {\em all} instances of line search (including those occurring later in the SCF process e.g. due to the gradient becoming too large again).
On average 7 Fock builds are needed before the gradient is sufficiently reduced to start quasi-Newton steps. This is consistent with the average of 3 line search iterations before stating L-BFGS because each line search takes two Fock builds and we need one initial Fock build.
These results are mostly an indication of the quality of the initial guess and the relative simplicity of the electronic structure in this test set.
The data in \cref{table:stats_overall} also illustrates the efficiency of the L-BFGS/TR combination used in QUOTR since almost half (8.8) of the total number of Fock builds (19.4) are spent in performing steps that ultimately use line searches.
The negligible difference between ``Line Search'' and ``Before L-BFGS'' statistics indicates that in most cases there is no need for line search after starting L-BFGS/TR steps.

\begin{table}
    \caption{QUOTR convergence statistics breakdown for G2-2/6-31G*.}
    \begin{tabular*}{0.48\textwidth}{@{\extracolsep{\fill}}clccc}
    \hline\hline
    {} & {} &  median &  max & mean \\
    \hline
    \multirow{3}{0.2cm}{$N_\mathrm{F}$} & Cumulative & 16 & 69 & 19.4 \\
    {} & Before L-BFGS & 7 & 21 & 7.8 \\
    {} & Line Search & 7 & 27 & 8.8 \\
    \hline
    \multirow{3}{0.2cm}{$N_\mathrm{I}$} & Cumulative & 12 & 55 & 14.2 \\
    {} & Before L-BFGS & 3 & 10 & 3.4 \\
    {} & Line Search & 3 & 13 & 3.9 \\ 
    \hline\hline
    \end{tabular*}
    \label{table:stats_overall}
\end{table}

\begin{figure}[h]
    \centering
    \includegraphics[width=3.25in]{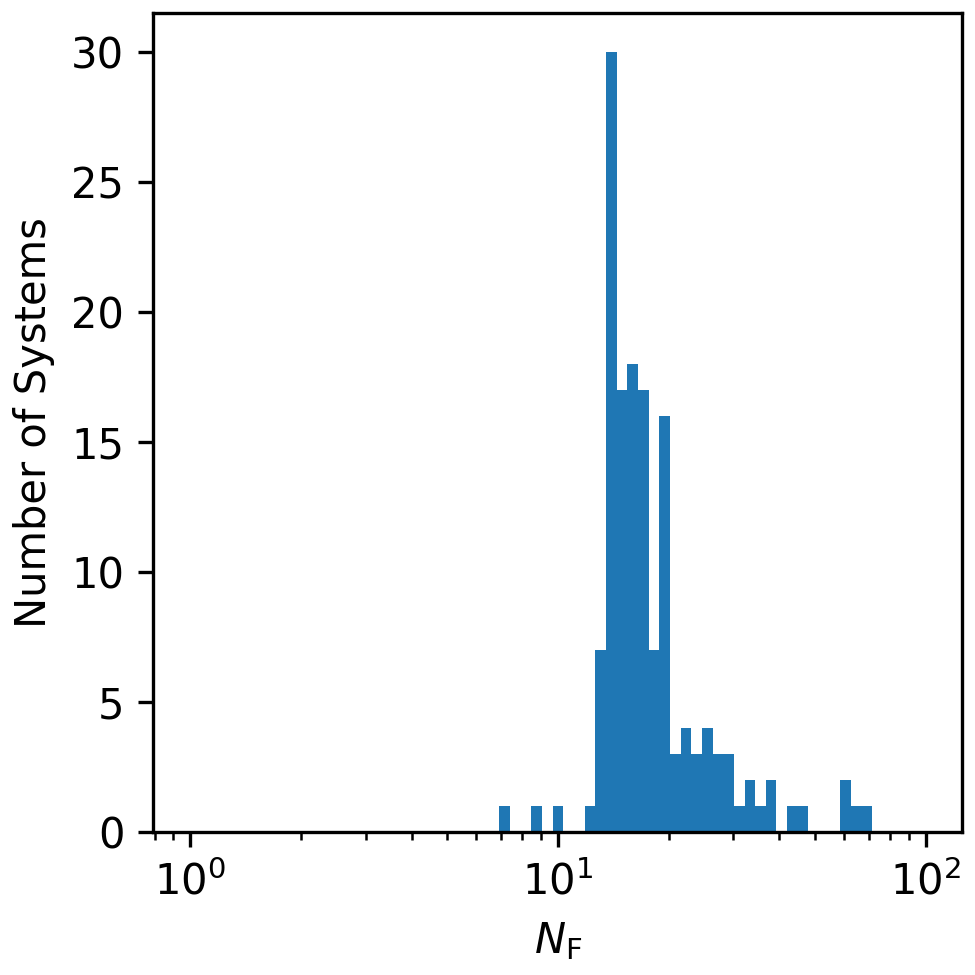}
    \caption{Histogram of QUOTR's $N_\mathrm{F}$ for G2-2/6-31G*.}
    \label{fig:nFock_all_hist_log}
\end{figure}

Although the G2-2 test set is composed of systems with relatively simple electronic structure, there is substantial variance within the set, as illustrated in \cref{fig:nFock_all_hist_log}. Although for most systems convergence is achieved in fewer than 30 Fock builds, there are more than 10 systems for which more Fock builds were required.

\begin{figure}[h]
    \centering
    \includegraphics[width=3.25 in]{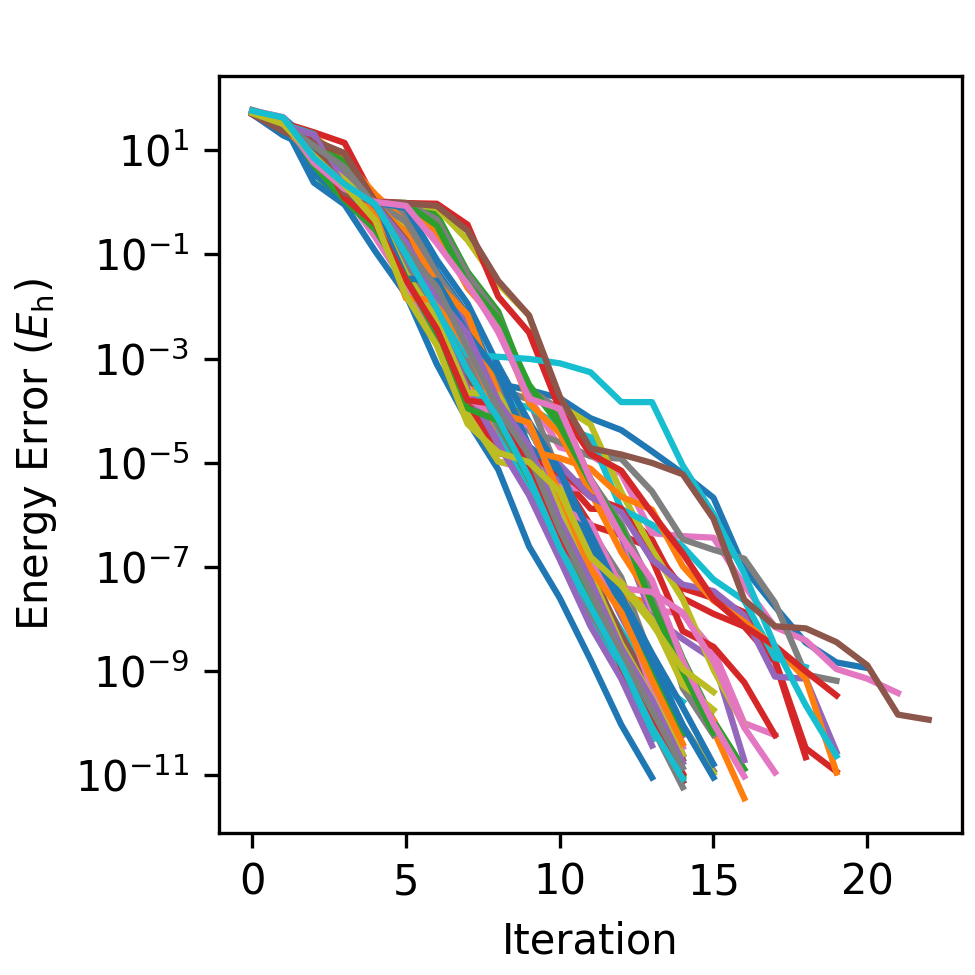}
    \caption{Convergence of AlCl$_3$ using QUOTR with random perturbation to extended H\"{u}ckel guess orbitals.}
    \label{fig:alcl3_50-rand}
\end{figure}

Quality of the initial guess unfortunately matters even with very robust solvers. Specifically, the need for random perturbation of the initial guess orbitals was found to be crucial for some systems with geometric symmetry.
In particular, we found for AlCl$_3$ that without breaking the symmetry of the extended H\"uckel orbitals, a local minimum at -1,619.598631 $E_{\mathrm{h}}$ was obtained by QUOTR.
However, when the perturbation was applied, a solution at -1,620.576010 $E_{\mathrm{h}}$ was consistently found.
To examine how the random perturbations to the initial guess impact convergence, we ran AlCl$_3$ with 50 different seeds for the random number generator.
The plot in \cref{fig:alcl3_50-rand} shows that when the minimum solution is accessible by symmetry, then QUOTR is robust in converging to the solution.

\subsection{Challenging Tests}
\label{subsection:difficult_systems}

We have now shown that QUOTR is competitive with standard RH/DIIS and competitive or superior to the representative quasi-Newton SCF solvers for the relatively easy test problems (G2 test set). 
To test the robustness of QUOTR we considered several prototypes of systems where the standard (RH/DIIS) usually fails outright, and even representative quasi-Newton heuristics struggle. We selected 3 types of problems where convergence difficulties often occur: (a) systems with small or vanishing HOMO-LUMO gap, (b) transition metal-containing systems, and (c) $f$-element containing systems.

\subsubsection{System with Vanishing HOMO-LUMO Gap.}
\label{subsection:vanishing_gap_system}

To demonstrate the performance of QUOTR for a more challenging problem, we considered a small neuropeptide  (1PLW\cite{VRG:marcotte:2004:BJ}) among several identified by Rudberg et al.\cite{VRG:rudberg:2012:JPCM},
for which the semilocal KS DFT SCF solutions could not be obtained using an RH-based solver.
We used QUOTR to obtain converged KS determinants with hybrid (PBE0 and B3LYP), semilocal (PBE), and local (LDA) functionals. 
The converged energies for the HOMO and LUMO along with the gap are displayed in \cref{table:1plw_analysis} using the 6-31G** orbital basis, with the def2-universal-J basis used for density fitting.
For the KS DFT calculations, QUOTR found somewhat lower energy solutions when the initial guess was not perturbed (about 0.36 m$E_{\mathrm{h}}$ for LDA, 0.31 m$E_{\mathrm{h}}$ for PBE, and 0.17 m$E_{\mathrm{h}}$ for B3LYP).
Thus, the data in \cref{table:1plw_analysis} is for the unperturbed initial guess.
As this system is significantly larger than the G2 tests, it is appropriate to comment on the orthogonality of the final orbitals.
We repeated the calculations for 1PLW and computed the orthogonality error, $||\mathbf{C}^{\dagger} \mathbf{SC} - \mathbf{1}||$, for the converged coefficient matrix. In all cases, this measure was on the order of $1 \times 10^{-12}$, indicating that the final solution does not deviate significantly from orthogonality.

\begin{table}
    \caption{Frontier orbital energies (eV) and the HOMO-LUMO gap for HF and KS DFT models of the 1PLW popyleptide (see text).}
    \begin{tabular*}{0.48\textwidth}{@{\extracolsep{\fill}}cccccc}
    \hline\hline
    {} & HF & LDA & PBE & PBE0 & B3LYP \\
    \hline
    HOMO & -6.38 & -3.26 & -2.53 & -2.70 & -2.75 \\
    LUMO & 0.86 & -3.25 & -2.52 & -2.34 & -2.47 \\
    Gap & 7.24 & 0.01 & 0.01 & 0.36 & 0.28 \\
    \hline\hline
    \end{tabular*}
    \label{table:1plw_analysis}
\end{table}

The sizeable 7.24 eV gap found for HF  nearly vanished with the hybrid DFT functionals (PBE0, B3LYP), with all values within 0.01 eV of the values found in Ref. \citenum{VRG:rudberg:2012:JPCM}.
For the LDA and PBE functionals, for which solutions could not be located in Ref. \citenum{VRG:rudberg:2012:JPCM}, QUOTR produced converged solutions with a nearly zero HOMO-LUMO gap! The origin of the vanishing gap in this and other similar biopolymers will be elaborated elsewhere, but we emphasize that the vanishing gap solution is the unphysical but ``correct'' solution, and QUOTR successfully located it. A key motivation for the development of QUOTR was the need to understand the origin of such unphysical solutions.

\begin{figure}[h]
    \centering
    \includegraphics[width=3.25in]{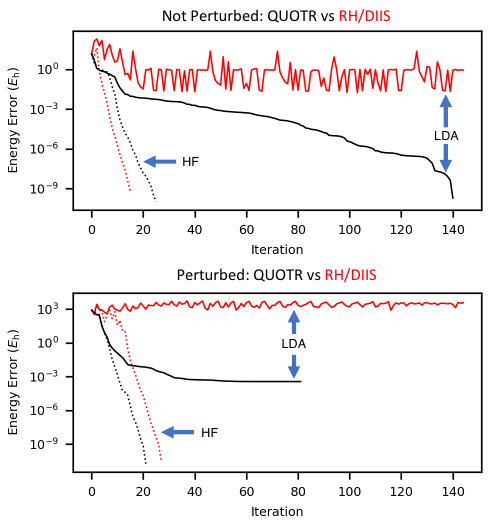}
    \caption{Convergence of HF and KS DFT for the 1PLW polypeptide (displayed energy error relative to the lowest energy obtained for each method).}
    \label{fig:1plw_converge}
\end{figure}

\cref{fig:1plw_converge} illustrates how the HF and KS LDA energies converge with QUOTR and RH/DIIS solvers.
The two panels in the Figure illustrate the impact of the starting orbitals on the solver convergence. While the ultimate outcome --- RH/DIIS did not converge for LDA, the rest of the combinations converged --- did not depend on whether the starting orbitals were perturbed or not, it took twice as many iterations for QUOTR to converge to the LDA solution without perturbation than with.
Note that the perturbation in these cases is applied to all orbitals, not just valence.
Both solvers converge at a similar rate for the HF case where the HOMO-LUMO gap is large, with approximately 20 or fewer iterations sufficient for microhartree accuracy.
With unperturbed guess the number of Fock builds for RH/DIIS and QUOTR are 17 and 43, respectively; with perturbed guess the corresponding counts are 28 and 34.

While QUOTR manages to locate the LDA solution correctly, its rate of convergence can be relatively slow. We identify regularization of the preconditioner as the likely culprit.
Since at the converged KS LDA solution the HOMO-LUMO gap is zero (see  \cref{table:1plw_analysis}) and the condition number of the exact Hessian is large and grows with the system size, the exact Hessian (hence, the preconditioner) can vary significantly as the solution is approached.
The last iteration where the preconditioner was recomputed was 51 (with non-perturbed guess) and 32 (with perturbed guess).

\subsubsection{Transition metal-containing molecules.}
\label{subsection:difficult_d_systems}

We considered two small systems that are well-known to be challenges to mean-field solvers: Cr$_2$ and CrC in their lowest-energy singlet states.\cite{VRG:daniels:2000:PCCP,VRG:helmich-paris:2021:JCP,VRG:dittmer:2023:JCP}

\cref{table:difficult_d_systems} reports the number of Fock builds necessary to converge HF and KS DFT using a variety of solvers. The first set of comparisons juxtaposes QUOTR (implemented in MPQC) against a quasi-Newton TRAH solver (implemented in the Orca program) and 2 variants of DIIS (both implemented in Orca). These computations used the core Hamiltonian initial guess throughout and the default convergence criteria of Orca: 5$\times 10^{-5}$ for the gradient norm, 1$\times 10^{-6} E_{\mathrm{h}}$ for the energy difference between iterations.
Although the core Hamiltonian initial guess is known to be poor, it was chosen to make sure that the same initial orbital set was used to bootstrap computations in MPQC and Orca. 
The choice of such a poor starting point makes the job of the orbital optimizer even more difficult.
It should be noted that TRAH uses a random number in one of the Davidson diagonalization start vectors which helps break symmetry, while for QUOTR we apply a small random unitary rotation to the initial guess (all orbitals, not just valence) with a maximum $\bm{\sigma}$ element of 0.01 for these systems.
Thus, the initial guess for QUOTR differs from the others by this perturbation, and the QUOTR initial guess is usually higher in energy (approx. 1 - 2 $E_{\mathrm{h}}$ for CrC and 4 - 4.5 $E_{\mathrm{h}}$ for Cr$_2$).
\begin{table}
    \caption{Performance of various SCF solvers for HF and KS DFT singlet ground states of CrC and Cr$_2$.$^a$}
    \begin{tabular*}{0.48\textwidth}{@{\extracolsep{\fill}}l|cccc|cc}
    \hline\hline
    {} & QUOTR & TRAH & DIIS$^b$ & KDIIS$^b$ & QUOTR$^c$ & CIAH$^c$ \\
    \hline
    CrC &  &  &  &  &  &  \\
    RHF & 162 & 377 & 44$^d$ & 47$^d$ & 205 & 164$^d$ \\
    LDA & 148 & 202 & \textemdash & 320$^d$ & 107 & 208 \\
    B3LYP & 129 & 300 & 25$^d$ & 440$^d$ & 91 & 240 \\
    \hline
    Cr$_2$ &  &  &  &  &  &  \\
    RHF & 249 & 295 & 26$^d$ & \textemdash & 472 & 160$^d$ \\
    LDA & 208 & 233 & 20 & 113$^d$ & 144 & 478$^d$ \\
    B3LYP & 123 & 267 & 18$^d$ & 201$^d$ & 169 & 330 \\
    \hline\hline
    \end{tabular*}
    
    \begin{tablenotes}
    \item $^a$ $N_\mathrm{F}$ are reported for each solver. The core Hamiltonian eigenstates were used as the initial guess, unless noted. The def2-TZVPP basis used throughout.
    \item $^b$ As implemented in ORCA.
    \item $^c$ initial guess: hcore + 1 Fock build and diagonalize
    \item $^d$ local minimum
    \end{tablenotes}

    \label{table:difficult_d_systems}

\end{table}

Comparing QUOTR to TRAH, we see that in all cases QUOTR requires fewer Fock builds.
The largest error in converged energies for QUOTR was for CrC with B3LYP, which was higher than TRAH by $1.1 \times 10^{-6} E_{\mathrm{h}}$.
This error is reasonable since the energies were only converged to $1 \times 10^{-6} E_{\mathrm{h}}$.

The results for DIIS and KDIIS look promising according to the number of Fock builds; however, local minima are very common, so the rapid convergence is deceiving. Only in the LDA case for Cr$_2$ was DIIS able to find a solution that is not a local minimum relative to QUOTR's solution.

The second batch of comparisons juxtaposes QUOTR (in MPQC) against the CIAH solver (in PySCF) using the custom variant of core Hamiltonian (hcore) guess in PySCF, namely the standard hcore guess followed up by a single RH iteration.
The first thing to notice is that QUOTR takes more Fock builds than CIAH for RHF, but fewer Fock builds for the other two methods.
However, CIAH converges to a local minimum for both systems when using RHF, indicating that QUOTR is more robust {\rm and/or} faster than CIAH in both cases.

\begin{figure}
    \caption{The valence orbitals viewed along the bond axis and the matching orbital energies (eV) for RHF and LDA singlet ground states of Cr$_2$.} 
    \label{fig:cr_orbs}
    \centering
    \includegraphics[width=3.25in]{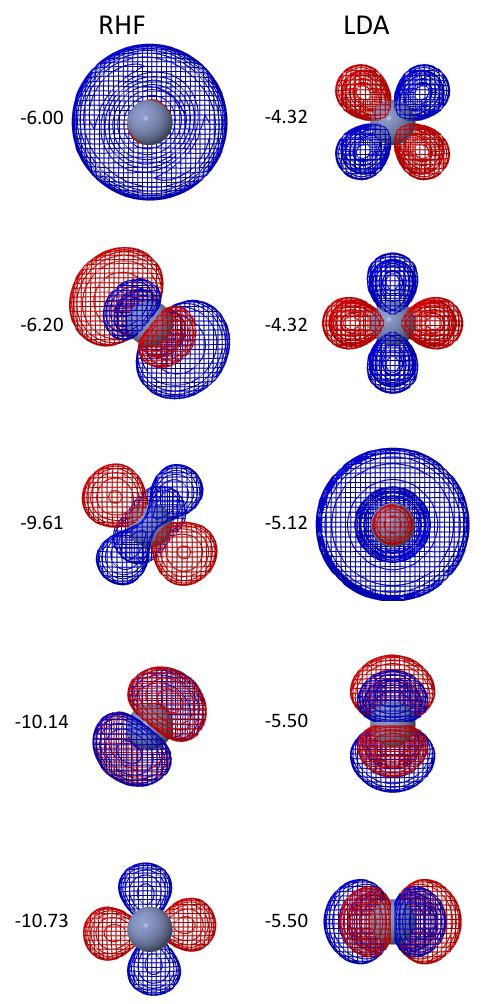}
\end{figure}

The very large number of Fock builds for Cr$_2$ with RHF merits further investigation.
While QUOTR does converge to the lowest energy solution that we could find, it does so at a cost of 249 Fock builds (or 472 for the comparison with PySCF hcore guess).
However, this difficulty is not  unique to QUOTR, as TRAH also takes nearly 300 Fock builds to achieve the same solution.
This is in contrast to the LDA solution, which seems to be generally easier to converge.
\cref{fig:cr_orbs} provides some insight into why RHF for Cr$_2$ is a difficult case.
Namely, the RHF solution located by QUOTR lacks cylindrical symmetry, in contrast to LDA.

To summarize: for CrC and Cr$_2$ QUOTR was able to locate the lowest-energy solution (unlike DIIS and CIAH) and was faster than TRAH.

\subsubsection{Actinide-containing molecule.}
\label{subsection:difficult_f_systems}

For the ultimate challenge, we considered the problem of converging the all-electron UHF orbitals in fermium mononitrate dication ($[\mathrm{Fm}(\mathrm{NO}_3)]^{2+}$), which is a known challenge for the SCF solver.\cite{VRG:penchoff:2018:AO}
Namely, Penchoff et al.\cite{VRG:penchoff:2018:AO} located 2 solutions, one with the expected +3 formal charge on Fm but located $\sim100$ kcal/mol {\em above} the correct ground state characterized by a +2 formal charge on Fm.

We used QUOTR with superposition of atomic densities guess orbitals, constructed as described in \cref{section:details}. Due to the complex optimization landscape in this system, it was necessary to explore the landscape of solutions by varying the initial guess. Thus the entire set of guess orbitals, or just their valence subset, was perturbed by pseudorandom unitaries generated using seven different integers (between 123 and 129) as the random engine seed. The results are displayed in \cref{table:fm_analysis}. 
The column labeled ``Energy Error'' is relative to the lowest energy solution that we obtained, which was for seed 2 with all orbitals perturbed (total energy -35,045.703224 $E_{\mathrm{h}}$).
\begin{table}
    \caption{Convergence statistics ($N_\mathrm{I}$,$N_\mathrm{F}$), energy error in kcal/mol ($\Delta E$), and atomic charges on the fermium atom (Q) for the UHF ground state of $[\mathrm{Fm}(\mathrm{NO}_3)]^{2+}$ obtained by QUOTR starting from a series of quasirandomly-perturbed minimal atomic guesses.}
    \begin{tabular*}{0.48\textwidth}{@{\extracolsep{\fill}}cccrcccrc}
    \hline\hline
    {} & \multicolumn{4}{c}{Valence Perturbed} & \multicolumn{4}{c}{All Perturbed} \\ \cmidrule(lr){2-5} \cmidrule(lr){6-9}
    Seed & $N_\mathrm{I}$ & $N_\mathrm{F}$ & \multicolumn{1}{c}{$\Delta E$} & Q & $N_\mathrm{I}$ & $N_\mathrm{F}$ & \multicolumn{1}{c}{$\Delta E$} & Q \\
    \hline
    1 & 124 & 189 & 40.19 & 2.70 & 131 & 201 & 0.58 & 1.81 \\
    2 & 132 & 199 & 40.13 & 2.70 & 102 & 168 & 0 & 1.81 \\
    3 & 203 & 274 & 0.76 & 1.81 & 205 & 295 & 0.54 & 1.81 \\
    4 & 136 & 191 & 40.16 & 2.70 & 225 & 307 & 0.04 & 1.81 \\
    5 & 189 & 262 & 40.16 & 2.70 & 209 & 290 & 0.25 & 1.81 \\
    6 & 167 & 229 & 40.34 & 2.70 & 144 & 215 & 0.34 & 1.81 \\
    7 & 126 & 170 & 0.92 & 1.82 & 155 & 243 & 0.35 & 1.81 \\
    \hline\hline
    \end{tabular*}
    \label{table:fm_analysis}
\end{table}
As can be seen, two types of solutions were found; the lower energy one has a formal +2 charge on Fm, and the other, roughly 40 kcal/mol higher in energy, has a +3 charge on Fm. The ground state energy agreed quite well with the value located by Penchoff et al. in Ref. \citenum{VRG:penchoff:2018:AO} using Molpro's RH/DIIS SCF solver using complicated guess obtained by merging converged fragment MOs (in fact, the ground state energy located by QUOTR is slightly lower in energy). Unfortunately, it was only possible to obtain $~10$ significant digits of precision in the energy, due to the impact of roundoff errors and the non-determinism of the Fock matrix construction in MPQC. As we did for 1PLW, we ran separate calculations to check the orthogonality of the converged orbitals. Again, we found errors on the order of $1 \times 10^{-12}$ in all cases. Clearly, all electron computations in heavy element systems with Gaussian AO bases that have high condition numbers will be increasingly untenable in double precision.

There are clearly many outstanding challenges suggested by the computational experiments on this actinide-containing complex. In particular, the sensitivity of the final solution to the initial guess suggests that various global (e.g., stochastic\cite{VRG:dittmer:2023:JCP}) approaches to the orbital optimization should be considered.

\begin{figure}[h]
    \centering
    \includegraphics[width=3.25in]{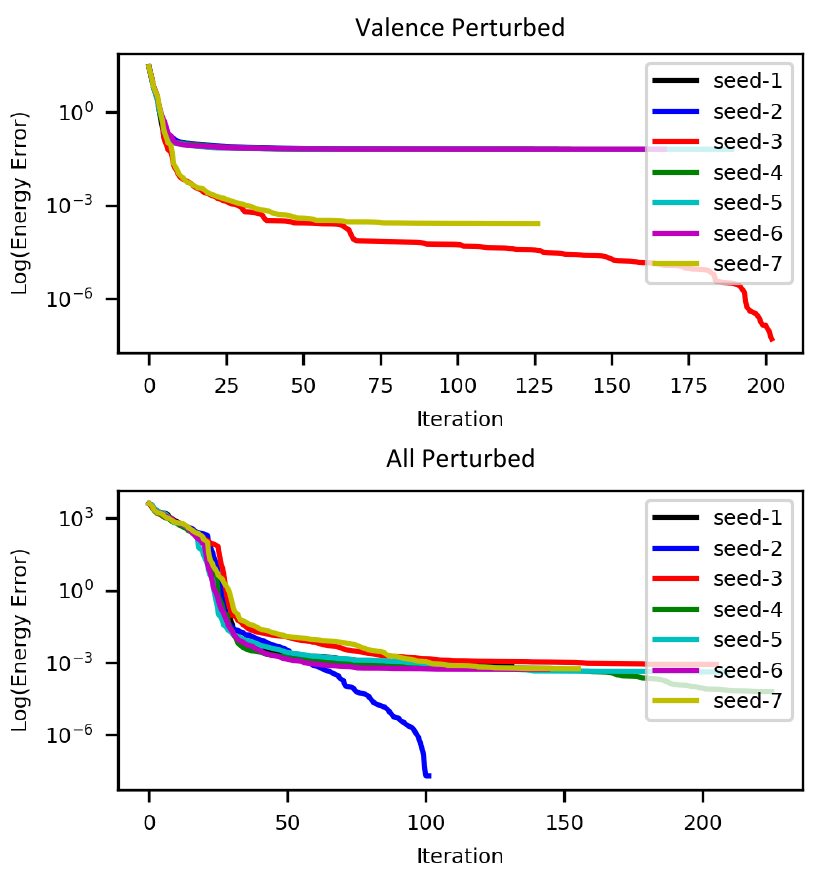}
    \caption{Convergence of X2C-UHF for $[\mathrm{Fm}(\mathrm{NO}_3)]^{2+}$ starting from a series of quasirandomly-perturbed minimal atomic guess orbitals; for each panel the energy error is defined relative to the lowest energy obtained in that panel's subset.}
    \label{fig:fmno3p2-valence-rand}
\end{figure}

\section{Summary}
\label{section:conclusion}

We have presented a state-of-the-art solver for quasi-Newton unitary optimization that combines the preconditioned L-BFGS orbital update with the trust-region step restriction method. The exploitation of the low-rank structure of the L-BFGS Hessian, including in solving the trust-region (sub)problem, makes the QUOTR solver remarkably efficient, approaching the efficiency of the mainstream RH/DIIS heuristics when applied to problems with easy optimization landscapes (like the standard G2 test set). When applied to problems with complex optimization landscapes (problems with vanishing HOMO-LUMO gaps, d- and f-element containing molecules) QUOTR matches or exceeds the robustness of representative quasi-Newton solvers, all at a significantly lower computational cost due to avoiding the exact Hessian evaluation.

While QUOTR guarantees convergence to a local stationary point, it is not able to guarantee global convergence due to the nonconvexity of the energy. However, its efficiency makes it a robust building block for even sophisticated solvers that combine efficient local minimum search with global (e.g., stochastic) landscape traversal.

%\section*{Author Contributions}
%We strongly encourage authors to include author contributions and recommend using \href{https://casrai.org/credit/}{CRediT} for standardised contribution descriptions. Please refer to our general \href{https://www.rsc.org/journals-books-databases/journal-authors-reviewers/author-responsibilities/}{author guidelines} for more information about authorship.

\section*{Conflicts of interest}
There are no conflicts to declare.

\section*{Acknowledgements}
This work was supported by the U.S. Department of Energy via award DE-SC0022327.

%%%END OF MAIN TEXT%%%


\balance


%%%REFERENCES%%%
\bibliography{vrgrefs, refs} %You need to replace "rsc" on this line with the name of your .bib file

\providecommand*{\mcitethebibliography}{\thebibliography}
\csname @ifundefined\endcsname{endmcitethebibliography}
{\let\endmcitethebibliography\endthebibliography}{}
\begin{mcitethebibliography}{112}
\providecommand*{\natexlab}[1]{#1}
\providecommand*{\mciteSetBstSublistMode}[1]{}
\providecommand*{\mciteSetBstMaxWidthForm}[2]{}
\providecommand*{\mciteBstWouldAddEndPuncttrue}
  {\def\EndOfBibitem{\unskip.}}
\providecommand*{\mciteBstWouldAddEndPunctfalse}
  {\let\EndOfBibitem\relax}
\providecommand*{\mciteSetBstMidEndSepPunct}[3]{}
\providecommand*{\mciteSetBstSublistLabelBeginEnd}[3]{}
\providecommand*{\EndOfBibitem}{}
\mciteSetBstSublistMode{f}
\mciteSetBstMaxWidthForm{subitem}
{(\emph{\alph{mcitesubitemcount}})}
\mciteSetBstSublistLabelBeginEnd{\mcitemaxwidthsubitemform\space}
{\relax}{\relax}

\bibitem[Hartree(1947)]{VRG:hartree:1947:RPP}
D.~R. Hartree, \emph{Rep. Prog. Phys.}, 1947, \textbf{11}, 113--143\relax
\mciteBstWouldAddEndPuncttrue
\mciteSetBstMidEndSepPunct{\mcitedefaultmidpunct}
{\mcitedefaultendpunct}{\mcitedefaultseppunct}\relax
\EndOfBibitem
\bibitem[Roothaan(1951)]{VRG:roothaan:1951:RMP}
C.~C.~J. Roothaan, \emph{Rev. Mod. Phys.}, 1951, \textbf{23}, 69--89\relax
\mciteBstWouldAddEndPuncttrue
\mciteSetBstMidEndSepPunct{\mcitedefaultmidpunct}
{\mcitedefaultendpunct}{\mcitedefaultseppunct}\relax
\EndOfBibitem
\bibitem[McWeeny(1956)]{VRG:mcweeny:1956:PRSLA}
R.~McWeeny, \emph{Proc. R. Soc. Lond. A}, 1956, \textbf{235}, 496--509\relax
\mciteBstWouldAddEndPuncttrue
\mciteSetBstMidEndSepPunct{\mcitedefaultmidpunct}
{\mcitedefaultendpunct}{\mcitedefaultseppunct}\relax
\EndOfBibitem
\bibitem[Levy and Berthier(1968)]{VRG:levy:1968:IJQC}
B.~Levy and G.~Berthier, \emph{Int. J. Quantum Chem.}, 1968, \textbf{2},
  307--319\relax
\mciteBstWouldAddEndPuncttrue
\mciteSetBstMidEndSepPunct{\mcitedefaultmidpunct}
{\mcitedefaultendpunct}{\mcitedefaultseppunct}\relax
\EndOfBibitem
\bibitem[Hillier and Saunders(1970)]{VRG:hillier:1970:PRSLA}
I.~H. Hillier and V.~R. Saunders, \emph{Proc. R. Soc. Lond. A}, 1970,
  \textbf{320}, 161--173\relax
\mciteBstWouldAddEndPuncttrue
\mciteSetBstMidEndSepPunct{\mcitedefaultmidpunct}
{\mcitedefaultendpunct}{\mcitedefaultseppunct}\relax
\EndOfBibitem
\bibitem[Hillier and Saunders(1970)]{VRG:hillier:1970:IJQC}
I.~H. Hillier and V.~R. Saunders, \emph{Int. J. Quantum Chem.}, 1970,
  \textbf{4}, 503--518\relax
\mciteBstWouldAddEndPuncttrue
\mciteSetBstMidEndSepPunct{\mcitedefaultmidpunct}
{\mcitedefaultendpunct}{\mcitedefaultseppunct}\relax
\EndOfBibitem
\bibitem[Levy(1973)]{VRG:levy:1973:CPL}
B.~Levy, \emph{Chem. Phys. Lett.}, 1973, \textbf{18}, 59--62\relax
\mciteBstWouldAddEndPuncttrue
\mciteSetBstMidEndSepPunct{\mcitedefaultmidpunct}
{\mcitedefaultendpunct}{\mcitedefaultseppunct}\relax
\EndOfBibitem
\bibitem[Seeger and Pople(1976)]{VRG:seeger:1976:JCP}
R.~Seeger and J.~A. Pople, \emph{J. Chem. Phys.}, 1976, \textbf{65},
  265--271\relax
\mciteBstWouldAddEndPuncttrue
\mciteSetBstMidEndSepPunct{\mcitedefaultmidpunct}
{\mcitedefaultendpunct}{\mcitedefaultseppunct}\relax
\EndOfBibitem
\bibitem[Douady \emph{et~al.}(1980)Douady, Ellinger, Subra, and
  Levy]{VRG:douady:1980:JCP}
J.~Douady, Y.~Ellinger, R.~Subra and B.~Levy, \emph{J. Chem. Phys.}, 1980,
  \textbf{72}, 1452--1462\relax
\mciteBstWouldAddEndPuncttrue
\mciteSetBstMidEndSepPunct{\mcitedefaultmidpunct}
{\mcitedefaultendpunct}{\mcitedefaultseppunct}\relax
\EndOfBibitem
\bibitem[Pulay(1980)]{VRG:pulay:1980:CPL}
P.~Pulay, \emph{Chem. Phys. Lett.}, 1980, \textbf{73}, 393--398\relax
\mciteBstWouldAddEndPuncttrue
\mciteSetBstMidEndSepPunct{\mcitedefaultmidpunct}
{\mcitedefaultendpunct}{\mcitedefaultseppunct}\relax
\EndOfBibitem
\bibitem[Bacskay(1981)]{VRG:bacskay:1981:CP}
G.~B. Bacskay, \emph{Chem. Phys.}, 1981, \textbf{61}, 385--404\relax
\mciteBstWouldAddEndPuncttrue
\mciteSetBstMidEndSepPunct{\mcitedefaultmidpunct}
{\mcitedefaultendpunct}{\mcitedefaultseppunct}\relax
\EndOfBibitem
\bibitem[Bacskay(1982)]{VRG:bacskay:1982:CP}
G.~B. Bacskay, \emph{Chem. Phys.}, 1982, \textbf{65}, 383--396\relax
\mciteBstWouldAddEndPuncttrue
\mciteSetBstMidEndSepPunct{\mcitedefaultmidpunct}
{\mcitedefaultendpunct}{\mcitedefaultseppunct}\relax
\EndOfBibitem
\bibitem[Pulay(1982)]{VRG:pulay:1982:JCC}
P.~Pulay, \emph{J. Comput. Chem.}, 1982, \textbf{3}, 556--560\relax
\mciteBstWouldAddEndPuncttrue
\mciteSetBstMidEndSepPunct{\mcitedefaultmidpunct}
{\mcitedefaultendpunct}{\mcitedefaultseppunct}\relax
\EndOfBibitem
\bibitem[{Head-Gordon} and Pople(1988)]{VRG:head-gordon:1988:JPC}
M.~{Head-Gordon} and J.~A. Pople, \emph{J. Phys. Chem.}, 1988, \textbf{92},
  3063--3069\relax
\mciteBstWouldAddEndPuncttrue
\mciteSetBstMidEndSepPunct{\mcitedefaultmidpunct}
{\mcitedefaultendpunct}{\mcitedefaultseppunct}\relax
\EndOfBibitem
\bibitem[Fischer and Almlof(1992)]{VRG:fischer:1992:JPC}
T.~H. Fischer and J.~Almlof, \emph{J. Phys. Chem.}, 1992, \textbf{96},
  9768--9774\relax
\mciteBstWouldAddEndPuncttrue
\mciteSetBstMidEndSepPunct{\mcitedefaultmidpunct}
{\mcitedefaultendpunct}{\mcitedefaultseppunct}\relax
\EndOfBibitem
\bibitem[Shepard(1993)]{VRG:shepard:1993:TCA}
R.~Shepard, \emph{Theoret. Chim. Acta}, 1993, \textbf{84}, 343--351\relax
\mciteBstWouldAddEndPuncttrue
\mciteSetBstMidEndSepPunct{\mcitedefaultmidpunct}
{\mcitedefaultendpunct}{\mcitedefaultseppunct}\relax
\EndOfBibitem
\bibitem[Rendell(1994)]{VRG:rendell:1994:CPL}
A.~P. Rendell, \emph{Chem. Phys. Lett.}, 1994, \textbf{229}, 204--210\relax
\mciteBstWouldAddEndPuncttrue
\mciteSetBstMidEndSepPunct{\mcitedefaultmidpunct}
{\mcitedefaultendpunct}{\mcitedefaultseppunct}\relax
\EndOfBibitem
\bibitem[Wong and Harrison(1995)]{VRG:wong:1995:JCC}
A.~T. Wong and R.~J. Harrison, \emph{J. Comput. Chem.}, 1995, \textbf{16},
  1291--1300\relax
\mciteBstWouldAddEndPuncttrue
\mciteSetBstMidEndSepPunct{\mcitedefaultmidpunct}
{\mcitedefaultendpunct}{\mcitedefaultseppunct}\relax
\EndOfBibitem
\bibitem[Chaban \emph{et~al.}(1997)Chaban, Schmidt, and
  Gordon]{VRG:chaban:1997:TCA}
G.~Chaban, M.~W. Schmidt and M.~S. Gordon, \emph{Theor. Chim. Acta}, 1997,
  \textbf{97}, 88--95\relax
\mciteBstWouldAddEndPuncttrue
\mciteSetBstMidEndSepPunct{\mcitedefaultmidpunct}
{\mcitedefaultendpunct}{\mcitedefaultseppunct}\relax
\EndOfBibitem
\bibitem[Daniels and Scuseria(2000)]{VRG:daniels:2000:PCCP}
A.~D. Daniels and G.~E. Scuseria, \emph{Phys. Chem. Chem. Phys.}, 2000,
  \textbf{2}, 2173--2176\relax
\mciteBstWouldAddEndPuncttrue
\mciteSetBstMidEndSepPunct{\mcitedefaultmidpunct}
{\mcitedefaultendpunct}{\mcitedefaultseppunct}\relax
\EndOfBibitem
\bibitem[Van~Voorhis and {Head-Gordon}(2002)]{VRG:vanvoorhis:2002:MP}
T.~Van~Voorhis and M.~{Head-Gordon}, \emph{Molecular Physics}, 2002,
  \textbf{100}, 1713--1721\relax
\mciteBstWouldAddEndPuncttrue
\mciteSetBstMidEndSepPunct{\mcitedefaultmidpunct}
{\mcitedefaultendpunct}{\mcitedefaultseppunct}\relax
\EndOfBibitem
\bibitem[VandeVondele and Hutter(2003)]{VRG:vandevondele:2003:JCP}
J.~VandeVondele and J.~Hutter, \emph{J. Chem. Phys.}, 2003, \textbf{118},
  4365--4369\relax
\mciteBstWouldAddEndPuncttrue
\mciteSetBstMidEndSepPunct{\mcitedefaultmidpunct}
{\mcitedefaultendpunct}{\mcitedefaultseppunct}\relax
\EndOfBibitem
\bibitem[Th{\o}gersen \emph{et~al.}(2004)Th{\o}gersen, Olsen, Yeager,
  J{\o}rgensen, Sa{\l}ek, and Helgaker]{VRG:thogersen:2004:JCP}
L.~Th{\o}gersen, J.~Olsen, D.~Yeager, P.~J{\o}rgensen, P.~Sa{\l}ek and
  T.~Helgaker, \emph{J. Chem. Phys.}, 2004, \textbf{121}, 16\relax
\mciteBstWouldAddEndPuncttrue
\mciteSetBstMidEndSepPunct{\mcitedefaultmidpunct}
{\mcitedefaultendpunct}{\mcitedefaultseppunct}\relax
\EndOfBibitem
\bibitem[Th{\o}gersen \emph{et~al.}(2005)Th{\o}gersen, Olsen, K{\"o}hn,
  J{\o}rgensen, Sa{\l}ek, and Helgaker]{VRG:thogersen:2005:JCP}
L.~Th{\o}gersen, J.~Olsen, A.~K{\"o}hn, P.~J{\o}rgensen, P.~Sa{\l}ek and
  T.~Helgaker, \emph{J. Chem. Phys.}, 2005, \textbf{123}, 074103\relax
\mciteBstWouldAddEndPuncttrue
\mciteSetBstMidEndSepPunct{\mcitedefaultmidpunct}
{\mcitedefaultendpunct}{\mcitedefaultseppunct}\relax
\EndOfBibitem
\bibitem[Yang \emph{et~al.}(2007)Yang, Meza, and Wang]{VRG:yang:2007:SJSC}
C.~Yang, J.~C. Meza and L.-W. Wang, \emph{SIAM J. Sci. Comput.}, 2007,
  \textbf{29}, 1854--1875\relax
\mciteBstWouldAddEndPuncttrue
\mciteSetBstMidEndSepPunct{\mcitedefaultmidpunct}
{\mcitedefaultendpunct}{\mcitedefaultseppunct}\relax
\EndOfBibitem
\bibitem[Sa{\l}ek \emph{et~al.}(2007)Sa{\l}ek, H{\o}st, Th{\o}gersen,
  J{\o}rgensen, Manninen, Olsen, Jans{\'i}k, Reine, Paw{\l}owski, Tellgren,
  Helgaker, and Coriani]{VRG:salek:2007:JCP}
P.~Sa{\l}ek, S.~H{\o}st, L.~Th{\o}gersen, P.~J{\o}rgensen, P.~Manninen,
  J.~Olsen, B.~Jans{\'i}k, S.~Reine, F.~Paw{\l}owski, E.~Tellgren, T.~Helgaker
  and S.~Coriani, \emph{J. Chem. Phys.}, 2007, \textbf{126}, 114110\relax
\mciteBstWouldAddEndPuncttrue
\mciteSetBstMidEndSepPunct{\mcitedefaultmidpunct}
{\mcitedefaultendpunct}{\mcitedefaultseppunct}\relax
\EndOfBibitem
\bibitem[Weber \emph{et~al.}(2008)Weber, VandeVondele, Hutter, and
  Niklasson]{VRG:weber:2008:JCP}
V.~Weber, J.~VandeVondele, J.~Hutter and A.~M.~N. Niklasson, \emph{J. Chem.
  Phys.}, 2008, \textbf{128}, 084113\relax
\mciteBstWouldAddEndPuncttrue
\mciteSetBstMidEndSepPunct{\mcitedefaultmidpunct}
{\mcitedefaultendpunct}{\mcitedefaultseppunct}\relax
\EndOfBibitem
\bibitem[H{\o}st \emph{et~al.}(2008)H{\o}st, Olsen, Jans{\'i}k, Th{\o}gersen,
  J{\o}rgensen, and Helgaker]{VRG:host:2008:JCP}
S.~H{\o}st, J.~Olsen, B.~Jans{\'i}k, L.~Th{\o}gersen, P.~J{\o}rgensen and
  T.~Helgaker, \emph{J. Chem. Phys.}, 2008, \textbf{129}, 124106\relax
\mciteBstWouldAddEndPuncttrue
\mciteSetBstMidEndSepPunct{\mcitedefaultmidpunct}
{\mcitedefaultendpunct}{\mcitedefaultseppunct}\relax
\EndOfBibitem
\bibitem[Baarman and VandeVondele(2011)]{VRG:baarman:2011:JCP}
K.~Baarman and J.~VandeVondele, \emph{J. Chem. Phys.}, 2011, \textbf{134},
  244104\relax
\mciteBstWouldAddEndPuncttrue
\mciteSetBstMidEndSepPunct{\mcitedefaultmidpunct}
{\mcitedefaultendpunct}{\mcitedefaultseppunct}\relax
\EndOfBibitem
\bibitem[Sun(2017)]{VRG:sun:2017:AP}
Q.~Sun, \emph{ArXiv161008423 Phys.}, 2017\relax
\mciteBstWouldAddEndPuncttrue
\mciteSetBstMidEndSepPunct{\mcitedefaultmidpunct}
{\mcitedefaultendpunct}{\mcitedefaultseppunct}\relax
\EndOfBibitem
\bibitem[{Helmich-Paris}(2021)]{VRG:helmich-paris:2021:JCP}
B.~{Helmich-Paris}, \emph{J. Chem. Phys.}, 2021, \textbf{154}, 164104\relax
\mciteBstWouldAddEndPuncttrue
\mciteSetBstMidEndSepPunct{\mcitedefaultmidpunct}
{\mcitedefaultendpunct}{\mcitedefaultseppunct}\relax
\EndOfBibitem
\bibitem[Nottoli \emph{et~al.}(2021)Nottoli, Gauss, and
  Lipparini]{VRG:nottoli:2021:MP}
T.~Nottoli, J.~Gauss and F.~Lipparini, \emph{Mol. Phys.}, 2021, \textbf{119},
  e1974590\relax
\mciteBstWouldAddEndPuncttrue
\mciteSetBstMidEndSepPunct{\mcitedefaultmidpunct}
{\mcitedefaultendpunct}{\mcitedefaultseppunct}\relax
\EndOfBibitem
\bibitem[Ivanov \emph{et~al.}(2021)Ivanov, J{\'o}nsson, Vegge, and
  J{\'o}nsson]{VRG:ivanov:2021:CPC}
A.~V. Ivanov, E.~{\"O}. J{\'o}nsson, T.~Vegge and H.~J{\'o}nsson, \emph{Comput.
  Phys. Commun.}, 2021, \textbf{267}, 108047\relax
\mciteBstWouldAddEndPuncttrue
\mciteSetBstMidEndSepPunct{\mcitedefaultmidpunct}
{\mcitedefaultendpunct}{\mcitedefaultseppunct}\relax
\EndOfBibitem
\bibitem[Seidl and Barca(2022)]{VRG:seidl:2022:JCTC}
C.~Seidl and G.~M.~J. Barca, \emph{J. Chem. Theory Comput.}, 2022, \textbf{18},
  4164--4176\relax
\mciteBstWouldAddEndPuncttrue
\mciteSetBstMidEndSepPunct{\mcitedefaultmidpunct}
{\mcitedefaultendpunct}{\mcitedefaultseppunct}\relax
\EndOfBibitem
\bibitem[Dittmer and Dreuw(2023)]{VRG:dittmer:2023:JCP}
L.~B. Dittmer and A.~Dreuw, \emph{J. Chem. Phys.}, 2023, \textbf{159},
  134104\relax
\mciteBstWouldAddEndPuncttrue
\mciteSetBstMidEndSepPunct{\mcitedefaultmidpunct}
{\mcitedefaultendpunct}{\mcitedefaultseppunct}\relax
\EndOfBibitem
\bibitem[Vavasis(1991)]{VRG:vavasis:1991:}
S.~Vavasis, \emph{Nonlinear Optimization: {{Complexity}} Issues}, {Oxford
  University Press}, 1991\relax
\mciteBstWouldAddEndPuncttrue
\mciteSetBstMidEndSepPunct{\mcitedefaultmidpunct}
{\mcitedefaultendpunct}{\mcitedefaultseppunct}\relax
\EndOfBibitem
\bibitem[Boumal \emph{et~al.}(2019)Boumal, Absil, and
  Cartis]{VRG:boumal:2019:IJNA}
N.~Boumal, P.-A. Absil and C.~Cartis, \emph{IMA J. Numer. Anal.}, 2019,
  \textbf{39}, 1--33\relax
\mciteBstWouldAddEndPuncttrue
\mciteSetBstMidEndSepPunct{\mcitedefaultmidpunct}
{\mcitedefaultendpunct}{\mcitedefaultseppunct}\relax
\EndOfBibitem
\bibitem[Hall(1951)]{VRG:hall:1951:PRSLAa}
G.~G. Hall, \emph{Proc. R. Soc. Lond. A}, 1951, \textbf{205}, 541--552\relax
\mciteBstWouldAddEndPuncttrue
\mciteSetBstMidEndSepPunct{\mcitedefaultmidpunct}
{\mcitedefaultendpunct}{\mcitedefaultseppunct}\relax
\EndOfBibitem
\bibitem[Anderson(1965)]{VRG:anderson:1965:JA}
D.~G. Anderson, \emph{J. ACM}, 1965, \textbf{12}, 547--560\relax
\mciteBstWouldAddEndPuncttrue
\mciteSetBstMidEndSepPunct{\mcitedefaultmidpunct}
{\mcitedefaultendpunct}{\mcitedefaultseppunct}\relax
\EndOfBibitem
\bibitem[Kudin \emph{et~al.}(2002)Kudin, Scuseria, and
  Canc{\`e}s]{VRG:kudin:2002:JCP}
K.~N. Kudin, G.~E. Scuseria and E.~Canc{\`e}s, \emph{J. Chem. Phys.}, 2002,
  \textbf{116}, 8255--8261\relax
\mciteBstWouldAddEndPuncttrue
\mciteSetBstMidEndSepPunct{\mcitedefaultmidpunct}
{\mcitedefaultendpunct}{\mcitedefaultseppunct}\relax
\EndOfBibitem
\bibitem[Garza and Scuseria(2012)]{VRG:garza:2012:JCP}
A.~J. Garza and G.~E. Scuseria, \emph{J. Chem. Phys.}, 2012, \textbf{137},
  054110\relax
\mciteBstWouldAddEndPuncttrue
\mciteSetBstMidEndSepPunct{\mcitedefaultmidpunct}
{\mcitedefaultendpunct}{\mcitedefaultseppunct}\relax
\EndOfBibitem
\bibitem[Harrison(2004)]{VRG:harrison:2004:JCC}
R.~J. Harrison, \emph{J. Comput. Chem.}, 2004, \textbf{25}, 328--334\relax
\mciteBstWouldAddEndPuncttrue
\mciteSetBstMidEndSepPunct{\mcitedefaultmidpunct}
{\mcitedefaultendpunct}{\mcitedefaultseppunct}\relax
\EndOfBibitem
\bibitem[Kollmar(1996)]{VRG:kollmar:1996:JCP}
C.~Kollmar, \emph{J. Chem. Phys.}, 1996, \textbf{105}, 8204--8212\relax
\mciteBstWouldAddEndPuncttrue
\mciteSetBstMidEndSepPunct{\mcitedefaultmidpunct}
{\mcitedefaultendpunct}{\mcitedefaultseppunct}\relax
\EndOfBibitem
\bibitem[Sun(2017)]{VRG:sun:2017:}
Q.~Sun, \emph{Co-Iterative Augmented {{Hessian}} Method for Orbital
  Optimization}, 2017\relax
\mciteBstWouldAddEndPuncttrue
\mciteSetBstMidEndSepPunct{\mcitedefaultmidpunct}
{\mcitedefaultendpunct}{\mcitedefaultseppunct}\relax
\EndOfBibitem
\bibitem[Rudberg(2012)]{VRG:rudberg:2012:JPCM}
E.~Rudberg, \emph{J. Phys.: Condens. Matter}, 2012, \textbf{24}, 072202\relax
\mciteBstWouldAddEndPuncttrue
\mciteSetBstMidEndSepPunct{\mcitedefaultmidpunct}
{\mcitedefaultendpunct}{\mcitedefaultseppunct}\relax
\EndOfBibitem
\bibitem[K{\"o}ppl and Werner(2016)]{VRG:koppl:2016:JCTC}
C.~K{\"o}ppl and H.-J. Werner, \emph{J. Chem. Theory Comput.}, 2016,
  \textbf{12}, 3122--3134\relax
\mciteBstWouldAddEndPuncttrue
\mciteSetBstMidEndSepPunct{\mcitedefaultmidpunct}
{\mcitedefaultendpunct}{\mcitedefaultseppunct}\relax
\EndOfBibitem
\bibitem[Lewis \emph{et~al.}(2016)Lewis, Calvin, and
  Valeev]{VRG:lewis:2016:JCTC}
C.~A. Lewis, J.~A. Calvin and E.~F. Valeev, \emph{J. Chem. Theory Comput.},
  2016, \textbf{12}, 5868--5880\relax
\mciteBstWouldAddEndPuncttrue
\mciteSetBstMidEndSepPunct{\mcitedefaultmidpunct}
{\mcitedefaultendpunct}{\mcitedefaultseppunct}\relax
\EndOfBibitem
\bibitem[Wang \emph{et~al.}(2020)Wang, Lewis, and Valeev]{VRG:wang:2020:JCP}
X.~Wang, C.~A. Lewis and E.~F. Valeev, \emph{J. Chem. Phys.}, 2020,
  \textbf{153}, 124116\relax
\mciteBstWouldAddEndPuncttrue
\mciteSetBstMidEndSepPunct{\mcitedefaultmidpunct}
{\mcitedefaultendpunct}{\mcitedefaultseppunct}\relax
\EndOfBibitem
\bibitem[Gilbert \emph{et~al.}(2008)Gilbert, Besley, and
  Gill]{VRG:gilbert:2008:JPCA}
A.~T.~B. Gilbert, N.~A. Besley and P.~M.~W. Gill, \emph{J. Phys. Chem. A},
  2008, \textbf{112}, 13164--13171\relax
\mciteBstWouldAddEndPuncttrue
\mciteSetBstMidEndSepPunct{\mcitedefaultmidpunct}
{\mcitedefaultendpunct}{\mcitedefaultseppunct}\relax
\EndOfBibitem
\bibitem[Stanton(1981)]{VRG:stanton:1981:JCP}
R.~E. Stanton, \emph{J. Chem. Phys.}, 1981, \textbf{75}, 3426--3432\relax
\mciteBstWouldAddEndPuncttrue
\mciteSetBstMidEndSepPunct{\mcitedefaultmidpunct}
{\mcitedefaultendpunct}{\mcitedefaultseppunct}\relax
\EndOfBibitem
\bibitem[Fletcher(1970)]{VRG:fletcher:1970:MP}
R.~Fletcher, \emph{Mol. Phys.}, 1970, \textbf{19}, 55--63\relax
\mciteBstWouldAddEndPuncttrue
\mciteSetBstMidEndSepPunct{\mcitedefaultmidpunct}
{\mcitedefaultendpunct}{\mcitedefaultseppunct}\relax
\EndOfBibitem
\bibitem[Chupin \emph{et~al.}(2021)Chupin, Dupuy, Legendre, and
  S{\'e}r{\'e}]{VRG:chupin:2021:EM}
M.~Chupin, M.-S. Dupuy, G.~Legendre and {\'E}.~S{\'e}r{\'e}, \emph{ESAIM:
  M2AN}, 2021, \textbf{55}, 2785--2825\relax
\mciteBstWouldAddEndPuncttrue
\mciteSetBstMidEndSepPunct{\mcitedefaultmidpunct}
{\mcitedefaultendpunct}{\mcitedefaultseppunct}\relax
\EndOfBibitem
\bibitem[Nocedal and Wright(2006)]{VRG:nocedal:2006:}
J.~Nocedal and S.~J. Wright, \emph{Numerical {{Optimization}}}, {Springer},
  {New York}, 2nd edn, 2006\relax
\mciteBstWouldAddEndPuncttrue
\mciteSetBstMidEndSepPunct{\mcitedefaultmidpunct}
{\mcitedefaultendpunct}{\mcitedefaultseppunct}\relax
\EndOfBibitem
\bibitem[Kollmar(1997)]{VRG:kollmar:1997:IJQC}
C.~Kollmar, \emph{Int. J. Quant. Chem.}, 1997, \textbf{62}, 617--637\relax
\mciteBstWouldAddEndPuncttrue
\mciteSetBstMidEndSepPunct{\mcitedefaultmidpunct}
{\mcitedefaultendpunct}{\mcitedefaultseppunct}\relax
\EndOfBibitem
\bibitem[Hu and Yang(2010)]{VRG:hu:2010:JCP}
X.~Hu and W.~Yang, \emph{J. Chem. Phys.}, 2010, \textbf{132}, 054109\relax
\mciteBstWouldAddEndPuncttrue
\mciteSetBstMidEndSepPunct{\mcitedefaultmidpunct}
{\mcitedefaultendpunct}{\mcitedefaultseppunct}\relax
\EndOfBibitem
\bibitem[Francisco \emph{et~al.}(2004)Francisco, Mart{\'{\i}}nez, and
  Mart{\'{\i}}nez]{VRG:francisco:2004:JCP}
J.~B. Francisco, J.~M. Mart{\'{\i}}nez and L.~Mart{\'{\i}}nez, \emph{J. Chem.
  Phys.}, 2004, \textbf{121}, 10863\relax
\mciteBstWouldAddEndPuncttrue
\mciteSetBstMidEndSepPunct{\mcitedefaultmidpunct}
{\mcitedefaultendpunct}{\mcitedefaultseppunct}\relax
\EndOfBibitem
\bibitem[Burdakov \emph{et~al.}(2017)Burdakov, Gong, Zikrin, and
  Yuan]{VRG:burdakov:2017:MPC}
O.~Burdakov, L.~Gong, S.~Zikrin and Y.-x. Yuan, \emph{Math. Prog. Comp.}, 2017,
  \textbf{9}, 101--134\relax
\mciteBstWouldAddEndPuncttrue
\mciteSetBstMidEndSepPunct{\mcitedefaultmidpunct}
{\mcitedefaultendpunct}{\mcitedefaultseppunct}\relax
\EndOfBibitem
\bibitem[Claxton and Smith(1971)]{VRG:claxton:1971:TCA}
T.~A. Claxton and N.~A. Smith, \emph{Theoret. Chim. Acta}, 1971, \textbf{22},
  399--402\relax
\mciteBstWouldAddEndPuncttrue
\mciteSetBstMidEndSepPunct{\mcitedefaultmidpunct}
{\mcitedefaultendpunct}{\mcitedefaultseppunct}\relax
\EndOfBibitem
\bibitem[Sleeman(1968)]{VRG:sleeman:1968:TCA}
D.~H. Sleeman, \emph{Theoret. Chim. Acta}, 1968, \textbf{11}, 135--144\relax
\mciteBstWouldAddEndPuncttrue
\mciteSetBstMidEndSepPunct{\mcitedefaultmidpunct}
{\mcitedefaultendpunct}{\mcitedefaultseppunct}\relax
\EndOfBibitem
\bibitem[Olsen and J{\o}rgensen(1982)]{VRG:olsen:1982:JCPa}
J.~Olsen and P.~J{\o}rgensen, \emph{J. Chem. Phys.}, 1982, \textbf{77},
  6109--6130\relax
\mciteBstWouldAddEndPuncttrue
\mciteSetBstMidEndSepPunct{\mcitedefaultmidpunct}
{\mcitedefaultendpunct}{\mcitedefaultseppunct}\relax
\EndOfBibitem
\bibitem[Kreplin \emph{et~al.}(2019)Kreplin, Knowles, and
  Werner]{VRG:kreplin:2019:JCP}
D.~A. Kreplin, P.~J. Knowles and H.-J. Werner, \emph{J. Chem. Phys.}, 2019,
  \textbf{150}, 194106\relax
\mciteBstWouldAddEndPuncttrue
\mciteSetBstMidEndSepPunct{\mcitedefaultmidpunct}
{\mcitedefaultendpunct}{\mcitedefaultseppunct}\relax
\EndOfBibitem
\bibitem[Yao and Umrigar(2021)]{VRG:yao:2021:JCTC}
Y.~Yao and C.~J. Umrigar, \emph{J. Chem. Theory Comput.}, 2021, \textbf{17},
  4183--4194\relax
\mciteBstWouldAddEndPuncttrue
\mciteSetBstMidEndSepPunct{\mcitedefaultmidpunct}
{\mcitedefaultendpunct}{\mcitedefaultseppunct}\relax
\EndOfBibitem
\bibitem[Jo{$\not$}rgensen and Simons(1983)]{VRG:jorgensen:1983:JCP}
P.~Jo{$\not$}rgensen and J.~Simons, \emph{J. Chem. Phys.}, 1983, \textbf{79},
  334--357\relax
\mciteBstWouldAddEndPuncttrue
\mciteSetBstMidEndSepPunct{\mcitedefaultmidpunct}
{\mcitedefaultendpunct}{\mcitedefaultseppunct}\relax
\EndOfBibitem
\bibitem[Jensen and J{\o}rgensen(1984)]{VRG:jensen:1984:JCP}
H.-J.~A. Jensen and P.~J{\o}rgensen, \emph{J. Chem. Phys.}, 1984, \textbf{80},
  1204--1214\relax
\mciteBstWouldAddEndPuncttrue
\mciteSetBstMidEndSepPunct{\mcitedefaultmidpunct}
{\mcitedefaultendpunct}{\mcitedefaultseppunct}\relax
\EndOfBibitem
\bibitem[Fletcher(1980)]{VRG:fletcher:1980:}
R.~Fletcher, \emph{Practical {{Methods}} of {{Optimization}}}, {Wiley}, {New
  York}, 1980, vol. Vol. 1\relax
\mciteBstWouldAddEndPuncttrue
\mciteSetBstMidEndSepPunct{\mcitedefaultmidpunct}
{\mcitedefaultendpunct}{\mcitedefaultseppunct}\relax
\EndOfBibitem
\bibitem[Thouless(1960)]{VRG:thouless:1960:NP}
D.~Thouless, \emph{Nuclear Physics}, 1960, \textbf{21}, 225--232\relax
\mciteBstWouldAddEndPuncttrue
\mciteSetBstMidEndSepPunct{\mcitedefaultmidpunct}
{\mcitedefaultendpunct}{\mcitedefaultseppunct}\relax
\EndOfBibitem
\bibitem[Levy(1969)]{VRG:levy:1969:CPL}
B.~Levy, \emph{Chem. Phys. Lett.}, 1969, \textbf{4}, 17--19\relax
\mciteBstWouldAddEndPuncttrue
\mciteSetBstMidEndSepPunct{\mcitedefaultmidpunct}
{\mcitedefaultendpunct}{\mcitedefaultseppunct}\relax
\EndOfBibitem
\bibitem[Helgaker \emph{et~al.}(2000)Helgaker, Jo{$\not$}rgensen, and
  Olsen]{VRG:helgaker:2000:}
T.~Helgaker, P.~Jo{$\not$}rgensen and J.~Olsen, \emph{Molecular
  Electronic-Structure Theory}, {John Wiley \& Sons, Ltd}, {Chichester, UK},
  1st edn, 2000\relax
\mciteBstWouldAddEndPuncttrue
\mciteSetBstMidEndSepPunct{\mcitedefaultmidpunct}
{\mcitedefaultendpunct}{\mcitedefaultseppunct}\relax
\EndOfBibitem
\bibitem[Moler and Van~Loan(1978)]{VRG:moler:1978:SR}
C.~Moler and C.~Van~Loan, \emph{SIAM Rev.}, 1978, \textbf{20}, 801--836\relax
\mciteBstWouldAddEndPuncttrue
\mciteSetBstMidEndSepPunct{\mcitedefaultmidpunct}
{\mcitedefaultendpunct}{\mcitedefaultseppunct}\relax
\EndOfBibitem
\bibitem[Byrd \emph{et~al.}(1994)Byrd, Nocedal, and
  Schnabel]{VRG:byrd:1994:MPa}
R.~H. Byrd, J.~Nocedal and R.~B. Schnabel, \emph{Math. Program.}, 1994,
  \textbf{63}, 129--156\relax
\mciteBstWouldAddEndPuncttrue
\mciteSetBstMidEndSepPunct{\mcitedefaultmidpunct}
{\mcitedefaultendpunct}{\mcitedefaultseppunct}\relax
\EndOfBibitem
\bibitem[L{\"o}wdin(1956)]{VRG:lowdin:1956:AiP}
P.-O. L{\"o}wdin, \emph{Advances in Physics}, 1956, \textbf{5}, 1--171\relax
\mciteBstWouldAddEndPuncttrue
\mciteSetBstMidEndSepPunct{\mcitedefaultmidpunct}
{\mcitedefaultendpunct}{\mcitedefaultseppunct}\relax
\EndOfBibitem
\bibitem[Abrudan \emph{et~al.}(2009)Abrudan, Eriksson, and
  Koivunen]{VRG:abrudan:2009:SP}
T.~Abrudan, J.~Eriksson and V.~Koivunen, \emph{Signal Processing}, 2009,
  \textbf{89}, 1704--1714\relax
\mciteBstWouldAddEndPuncttrue
\mciteSetBstMidEndSepPunct{\mcitedefaultmidpunct}
{\mcitedefaultendpunct}{\mcitedefaultseppunct}\relax
\EndOfBibitem
\bibitem[Hoffmann(1963)]{VRG:hoffmann:1963:JCP}
R.~Hoffmann, \emph{J. Chem. Phys.}, 1963, \textbf{39}, 1397--1412\relax
\mciteBstWouldAddEndPuncttrue
\mciteSetBstMidEndSepPunct{\mcitedefaultmidpunct}
{\mcitedefaultendpunct}{\mcitedefaultseppunct}\relax
\EndOfBibitem
\bibitem[Ammeter \emph{et~al.}(1978)Ammeter, Buergi, Thibeault, and
  Hoffmann]{VRG:ammeter:1978:JACS}
J.~H. Ammeter, H.~B. Buergi, J.~C. Thibeault and R.~Hoffmann, \emph{J. Am.
  Chem. Soc.}, 1978, \textbf{100}, 3686--3692\relax
\mciteBstWouldAddEndPuncttrue
\mciteSetBstMidEndSepPunct{\mcitedefaultmidpunct}
{\mcitedefaultendpunct}{\mcitedefaultseppunct}\relax
\EndOfBibitem
\bibitem[Lehtola(2019)]{VRG:lehtola:2019:JCTC}
S.~Lehtola, \emph{J. Chem. Theory Comput.}, 2019, \textbf{15}, 1593--1604\relax
\mciteBstWouldAddEndPuncttrue
\mciteSetBstMidEndSepPunct{\mcitedefaultmidpunct}
{\mcitedefaultendpunct}{\mcitedefaultseppunct}\relax
\EndOfBibitem
\bibitem[Norman and Jensen(2012)]{VRG:norman:2012:CPL}
P.~Norman and H.~J.~A. Jensen, \emph{Chem. Phys. Lett.}, 2012, \textbf{531},
  229--235\relax
\mciteBstWouldAddEndPuncttrue
\mciteSetBstMidEndSepPunct{\mcitedefaultmidpunct}
{\mcitedefaultendpunct}{\mcitedefaultseppunct}\relax
\EndOfBibitem
\bibitem[Fischer(1972)]{VRG:fischer:1972:ADaNDT}
C.~F. Fischer, \emph{Atomic Data and Nuclear Data Tables}, 1972, \textbf{4},
  301--399\relax
\mciteBstWouldAddEndPuncttrue
\mciteSetBstMidEndSepPunct{\mcitedefaultmidpunct}
{\mcitedefaultendpunct}{\mcitedefaultseppunct}\relax
\EndOfBibitem
\bibitem[Peng \emph{et~al.}(2020)Peng, Lewis, Wang, Clement, Pierce, Rishi,
  Pavo{\v s}evi{\'c}, Slattery, Zhang, Teke, Kumar, Masteran, Asadchev, Calvin,
  and Valeev]{VRG:peng:2020:JCP}
C.~Peng, C.~A. Lewis, X.~Wang, M.~C. Clement, K.~Pierce, V.~Rishi, F.~Pavo{\v
  s}evi{\'c}, S.~Slattery, J.~Zhang, N.~Teke, A.~Kumar, C.~Masteran,
  A.~Asadchev, J.~A. Calvin and E.~F. Valeev, \emph{J. Chem. Phys.}, 2020,
  \textbf{153}, 044120\relax
\mciteBstWouldAddEndPuncttrue
\mciteSetBstMidEndSepPunct{\mcitedefaultmidpunct}
{\mcitedefaultendpunct}{\mcitedefaultseppunct}\relax
\EndOfBibitem
\bibitem[Ditchfield \emph{et~al.}(1971)Ditchfield, Hehre, and
  Pople]{VRG:ditchfield:1971:JCP}
R.~Ditchfield, W.~J. Hehre and J.~A. Pople, \emph{J. Chem. Phys.}, 1971,
  \textbf{54}, 724--728\relax
\mciteBstWouldAddEndPuncttrue
\mciteSetBstMidEndSepPunct{\mcitedefaultmidpunct}
{\mcitedefaultendpunct}{\mcitedefaultseppunct}\relax
\EndOfBibitem
\bibitem[Hehre \emph{et~al.}(1972)Hehre, Ditchfield, and
  Pople]{VRG:hehre:1972:JCP}
W.~J. Hehre, R.~Ditchfield and J.~A. Pople, \emph{J. Chem. Phys.}, 1972,
  \textbf{56}, 2257--2261\relax
\mciteBstWouldAddEndPuncttrue
\mciteSetBstMidEndSepPunct{\mcitedefaultmidpunct}
{\mcitedefaultendpunct}{\mcitedefaultseppunct}\relax
\EndOfBibitem
\bibitem[Hariharan and Pople(1973)]{VRG:hariharan:1973:TCA}
P.~C. Hariharan and J.~A. Pople, \emph{Theoret. Chim. Acta}, 1973, \textbf{28},
  213--222\relax
\mciteBstWouldAddEndPuncttrue
\mciteSetBstMidEndSepPunct{\mcitedefaultmidpunct}
{\mcitedefaultendpunct}{\mcitedefaultseppunct}\relax
\EndOfBibitem
\bibitem[Dill and Pople(1975)]{VRG:dill:1975:JCP}
J.~D. Dill and J.~A. Pople, \emph{J. Chem. Phys.}, 1975, \textbf{62},
  2921--2923\relax
\mciteBstWouldAddEndPuncttrue
\mciteSetBstMidEndSepPunct{\mcitedefaultmidpunct}
{\mcitedefaultendpunct}{\mcitedefaultseppunct}\relax
\EndOfBibitem
\bibitem[Binkley and Pople(1977)]{VRG:binkley:1977:JCP}
J.~S. Binkley and J.~A. Pople, \emph{J. Chem. Phys.}, 1977, \textbf{66},
  879--880\relax
\mciteBstWouldAddEndPuncttrue
\mciteSetBstMidEndSepPunct{\mcitedefaultmidpunct}
{\mcitedefaultendpunct}{\mcitedefaultseppunct}\relax
\EndOfBibitem
\bibitem[Gordon \emph{et~al.}(1982)Gordon, Binkley, Pople, Pietro, and
  Hehre]{VRG:gordon:1982:JACS}
M.~S. Gordon, J.~S. Binkley, J.~A. Pople, W.~J. Pietro and W.~J. Hehre,
  \emph{J. Am. Chem. Soc.}, 1982, \textbf{104}, 2797--2803\relax
\mciteBstWouldAddEndPuncttrue
\mciteSetBstMidEndSepPunct{\mcitedefaultmidpunct}
{\mcitedefaultendpunct}{\mcitedefaultseppunct}\relax
\EndOfBibitem
\bibitem[Francl \emph{et~al.}(1982)Francl, Pietro, Hehre, Binkley, Gordon,
  DeFrees, and Pople]{VRG:francl:1982:JCP}
M.~M. Francl, W.~J. Pietro, W.~J. Hehre, J.~S. Binkley, M.~S. Gordon, D.~J.
  DeFrees and J.~A. Pople, \emph{J. Chem. Phys.}, 1982, \textbf{77},
  3654--3665\relax
\mciteBstWouldAddEndPuncttrue
\mciteSetBstMidEndSepPunct{\mcitedefaultmidpunct}
{\mcitedefaultendpunct}{\mcitedefaultseppunct}\relax
\EndOfBibitem
\bibitem[Krishnan \emph{et~al.}(1980)Krishnan, Binkley, Seeger, and
  Pople]{VRG:krishnan:1980:JCP}
R.~Krishnan, J.~S. Binkley, R.~Seeger and J.~A. Pople, \emph{J. Chem. Phys.},
  1980, \textbf{72}, 650--654\relax
\mciteBstWouldAddEndPuncttrue
\mciteSetBstMidEndSepPunct{\mcitedefaultmidpunct}
{\mcitedefaultendpunct}{\mcitedefaultseppunct}\relax
\EndOfBibitem
\bibitem[McLean and Chandler(1980)]{VRG:mclean:1980:JCP}
A.~D. McLean and G.~S. Chandler, \emph{J. Chem. Phys.}, 1980, \textbf{72},
  5639--5648\relax
\mciteBstWouldAddEndPuncttrue
\mciteSetBstMidEndSepPunct{\mcitedefaultmidpunct}
{\mcitedefaultendpunct}{\mcitedefaultseppunct}\relax
\EndOfBibitem
\bibitem[Clark \emph{et~al.}(1983)Clark, Chandrasekhar, Spitznagel, and
  Schleyer]{VRG:clark:1983:JCC}
T.~Clark, J.~Chandrasekhar, G.~W. Spitznagel and P.~V.~R. Schleyer, \emph{J.
  Comput. Chem.}, 1983, \textbf{4}, 294--301\relax
\mciteBstWouldAddEndPuncttrue
\mciteSetBstMidEndSepPunct{\mcitedefaultmidpunct}
{\mcitedefaultendpunct}{\mcitedefaultseppunct}\relax
\EndOfBibitem
\bibitem[Spitznagel \emph{et~al.}(1987)Spitznagel, Clark, {von Ragu{\'e}
  Schleyer}, and Hehre]{VRG:spitznagel:1987:JCC}
G.~W. Spitznagel, T.~Clark, P.~{von Ragu{\'e} Schleyer} and W.~J. Hehre,
  \emph{J. Comput. Chem.}, 1987, \textbf{8}, 1109--1116\relax
\mciteBstWouldAddEndPuncttrue
\mciteSetBstMidEndSepPunct{\mcitedefaultmidpunct}
{\mcitedefaultendpunct}{\mcitedefaultseppunct}\relax
\EndOfBibitem
\bibitem[Weigend and Ahlrichs(2005)]{VRG:weigend:2005:PCCP}
F.~Weigend and R.~Ahlrichs, \emph{Phys. Chem. Chem. Phys.}, 2005, \textbf{7},
  3297\relax
\mciteBstWouldAddEndPuncttrue
\mciteSetBstMidEndSepPunct{\mcitedefaultmidpunct}
{\mcitedefaultendpunct}{\mcitedefaultseppunct}\relax
\EndOfBibitem
\bibitem[{de Jong} \emph{et~al.}(2001){de Jong}, Harrison, and
  Dixon]{VRG:dejong:2001:JCP}
W.~A. {de Jong}, R.~J. Harrison and D.~A. Dixon, \emph{J. Chem. Phys.}, 2001,
  \textbf{114}, 48\relax
\mciteBstWouldAddEndPuncttrue
\mciteSetBstMidEndSepPunct{\mcitedefaultmidpunct}
{\mcitedefaultendpunct}{\mcitedefaultseppunct}\relax
\EndOfBibitem
\bibitem[Feng and Peterson(2017)]{VRG:feng:2017:JCP}
R.~Feng and K.~A. Peterson, \emph{J. Chem. Phys.}, 2017, \textbf{147},
  084108\relax
\mciteBstWouldAddEndPuncttrue
\mciteSetBstMidEndSepPunct{\mcitedefaultmidpunct}
{\mcitedefaultendpunct}{\mcitedefaultseppunct}\relax
\EndOfBibitem
\bibitem[Weigend(2006)]{VRG:weigend:2006:PCCP}
F.~Weigend, \emph{Phys. Chem. Chem. Phys.}, 2006, \textbf{8}, 1057\relax
\mciteBstWouldAddEndPuncttrue
\mciteSetBstMidEndSepPunct{\mcitedefaultmidpunct}
{\mcitedefaultendpunct}{\mcitedefaultseppunct}\relax
\EndOfBibitem
\bibitem[Schuchardt \emph{et~al.}(2007)Schuchardt, Didier, Elsethagen, Sun,
  Gurumoorthi, Chase, Li, and Windus]{VRG:schuchardt:2007:JCIM}
K.~L. Schuchardt, B.~T. Didier, T.~Elsethagen, L.~Sun, V.~Gurumoorthi,
  J.~Chase, J.~Li and T.~L. Windus, \emph{J. Chem. Inf. Model.}, 2007,
  \textbf{47}, 1045--1052\relax
\mciteBstWouldAddEndPuncttrue
\mciteSetBstMidEndSepPunct{\mcitedefaultmidpunct}
{\mcitedefaultendpunct}{\mcitedefaultseppunct}\relax
\EndOfBibitem
\bibitem[Tsuchiya \emph{et~al.}(2001)Tsuchiya, Abe, Nakajima, and
  Hirao]{VRG:tsuchiya:2001:JCP}
T.~Tsuchiya, M.~Abe, T.~Nakajima and K.~Hirao, \emph{J. Chem. Phys.}, 2001,
  \textbf{115}, 4463--4472\relax
\mciteBstWouldAddEndPuncttrue
\mciteSetBstMidEndSepPunct{\mcitedefaultmidpunct}
{\mcitedefaultendpunct}{\mcitedefaultseppunct}\relax
\EndOfBibitem
\bibitem[Clement \emph{et~al.}(2021)Clement, Wang, and
  Valeev]{VRG:clement:2021:JCTC}
M.~C. Clement, X.~Wang and E.~F. Valeev, \emph{J. Chem. Theory Comput.}, 2021,
  \textbf{17}, 7406--7415\relax
\mciteBstWouldAddEndPuncttrue
\mciteSetBstMidEndSepPunct{\mcitedefaultmidpunct}
{\mcitedefaultendpunct}{\mcitedefaultseppunct}\relax
\EndOfBibitem
\bibitem[Curtiss \emph{et~al.}(1997)Curtiss, Raghavachari, Redfern, and
  Pople]{VRG:curtiss:1997:JCP}
L.~A. Curtiss, K.~Raghavachari, P.~C. Redfern and J.~A. Pople, \emph{J. Chem.
  Phys.}, 1997, \textbf{106}, 1063--1079\relax
\mciteBstWouldAddEndPuncttrue
\mciteSetBstMidEndSepPunct{\mcitedefaultmidpunct}
{\mcitedefaultendpunct}{\mcitedefaultseppunct}\relax
\EndOfBibitem
\bibitem[Johnson(2002)]{VRG:johnson:2002:}
{\relax RD}.~Johnson, \emph{Computational {{Chemistry Comparison}} and
  {{Benchmark Database}}, {{NIST Standard Reference Database}} 101}, 2002\relax
\mciteBstWouldAddEndPuncttrue
\mciteSetBstMidEndSepPunct{\mcitedefaultmidpunct}
{\mcitedefaultendpunct}{\mcitedefaultseppunct}\relax
\EndOfBibitem
\bibitem[Frisch \emph{et~al.}(2013)Frisch, Trucks, Schlegel, Scuseria, Robb,
  Cheeseman, Scalmani, Barone, Mennucci, Petersson, Nakatsuji, Caricato, Li,
  Hratchian, Izmaylov, Bloino, Zheng, Sonnenberg, Hada, Ehara, Toyota, Fukuda,
  Hasegawa, Ishida, Nakajima, Honda, Kitao, Nakai, Vreven, Montgomery, Peralta,
  Ogliaro, Bearpark, Heyd, Brothers, Kudin, Staroverov, Keith, Kobayashi,
  Normand, Raghavachari, Rendell, Burant, Iyengar, Tomasi, Cossi, Rega, Millam,
  Klene, Knox, Cross, Bakken, Adamo, Jaramillo, Gomperts, Stratmann, Yazyev,
  Austin, Cammi, Pomelli, Ochterski, Martin, Morokuma, Zakrzewski, Voth,
  Salvador, Dannenberg, Dapprich, Daniels, Farkas, Foresman, Ortiz, Cioslowski,
  and Fox]{VRG:frisch:2013:}
M.~J. Frisch, G.~W. Trucks, H.~B. Schlegel, G.~E. Scuseria, M.~A. Robb, J.~R.
  Cheeseman, G.~Scalmani, V.~Barone, B.~Mennucci, G.~A. Petersson,
  H.~Nakatsuji, M.~Caricato, X.~Li, H.~P. Hratchian, A.~F. Izmaylov, J.~Bloino,
  G.~Zheng, J.~L. Sonnenberg, M.~Hada, M.~Ehara, K.~Toyota, R.~Fukuda,
  J.~Hasegawa, M.~Ishida, T.~Nakajima, Y.~Honda, O.~Kitao, H.~Nakai, T.~Vreven,
  J.~A.~{\relax Jr}. Montgomery, J.~E. Peralta, F.~Ogliaro, M.~Bearpark, J.~J.
  Heyd, E.~Brothers, K.~N. Kudin, V.~N. Staroverov, T.~Keith, R.~Kobayashi,
  J.~Normand, K.~Raghavachari, A.~Rendell, J.~C. Burant, S.~S. Iyengar,
  J.~Tomasi, M.~Cossi, N.~Rega, J.~M. Millam, M.~Klene, J.~E. Knox, J.~B.
  Cross, V.~Bakken, C.~Adamo, J.~Jaramillo, R.~Gomperts, R.~E. Stratmann,
  O.~Yazyev, A.~J. Austin, R.~Cammi, C.~Pomelli, J.~W. Ochterski, R.~L. Martin,
  K.~Morokuma, V.~G. Zakrzewski, G.~A. Voth, P.~Salvador, J.~J. Dannenberg,
  S.~Dapprich, A.~D. Daniels, O.~Farkas, J.~B. Foresman, J.~V. Ortiz,
  J.~Cioslowski and D.~J. Fox, \emph{Gaussian 09}, Gaussian, Inc., 2013\relax
\mciteBstWouldAddEndPuncttrue
\mciteSetBstMidEndSepPunct{\mcitedefaultmidpunct}
{\mcitedefaultendpunct}{\mcitedefaultseppunct}\relax
\EndOfBibitem
\bibitem[Petrone \emph{et~al.}(2018)Petrone, {Williams-Young}, Sun, Stetina,
  and Li]{VRG:petrone:2018:EPJB}
A.~Petrone, D.~B. {Williams-Young}, S.~Sun, T.~F. Stetina and X.~Li, \emph{Eur.
  Phys. J. B}, 2018, \textbf{91}, 169\relax
\mciteBstWouldAddEndPuncttrue
\mciteSetBstMidEndSepPunct{\mcitedefaultmidpunct}
{\mcitedefaultendpunct}{\mcitedefaultseppunct}\relax
\EndOfBibitem
\bibitem[Lehtola \emph{et~al.}(2018)Lehtola, Steigemann, Oliveira, and
  Marques]{VRG:lehtola:2018:S}
S.~Lehtola, C.~Steigemann, M.~J. Oliveira and M.~A. Marques, \emph{SoftwareX},
  2018, \textbf{7}, 1--5\relax
\mciteBstWouldAddEndPuncttrue
\mciteSetBstMidEndSepPunct{\mcitedefaultmidpunct}
{\mcitedefaultendpunct}{\mcitedefaultseppunct}\relax
\EndOfBibitem
\bibitem[Mura and Knowles(1996)]{VRG:mura:1996:JCP}
M.~E. Mura and P.~J. Knowles, \emph{J. Chem. Phys.}, 1996, \textbf{104},
  9848--9858\relax
\mciteBstWouldAddEndPuncttrue
\mciteSetBstMidEndSepPunct{\mcitedefaultmidpunct}
{\mcitedefaultendpunct}{\mcitedefaultseppunct}\relax
\EndOfBibitem
\bibitem[Lebedev and Laikov(1999)]{VRG:lebedev:1999:DM}
V.~I. Lebedev and D.~N. Laikov, \emph{Dokl. Math.}, 1999, \textbf{59},
  477--481\relax
\mciteBstWouldAddEndPuncttrue
\mciteSetBstMidEndSepPunct{\mcitedefaultmidpunct}
{\mcitedefaultendpunct}{\mcitedefaultseppunct}\relax
\EndOfBibitem
\bibitem[Dirac(1930)]{VRG:dirac:1930:MPCPS}
P.~A.~M. Dirac, \emph{Math. Proc. Camb. Phil. Soc.}, 1930, \textbf{26},
  376--385\relax
\mciteBstWouldAddEndPuncttrue
\mciteSetBstMidEndSepPunct{\mcitedefaultmidpunct}
{\mcitedefaultendpunct}{\mcitedefaultseppunct}\relax
\EndOfBibitem
\bibitem[Vosko \emph{et~al.}(1980)Vosko, Wilk, and Nusair]{VRG:vosko:1980:CJP}
S.~H. Vosko, L.~Wilk and M.~Nusair, \emph{Can. J. Phys.}, 1980, \textbf{58},
  1200--1211\relax
\mciteBstWouldAddEndPuncttrue
\mciteSetBstMidEndSepPunct{\mcitedefaultmidpunct}
{\mcitedefaultendpunct}{\mcitedefaultseppunct}\relax
\EndOfBibitem
\bibitem[Marcotte \emph{et~al.}(2004)Marcotte, Separovic, Auger, and
  Gagn{\'e}]{VRG:marcotte:2004:BJ}
I.~Marcotte, F.~Separovic, M.~Auger and S.~M. Gagn{\'e}, \emph{Biophys. J.},
  2004, \textbf{86}, 1587--1600\relax
\mciteBstWouldAddEndPuncttrue
\mciteSetBstMidEndSepPunct{\mcitedefaultmidpunct}
{\mcitedefaultendpunct}{\mcitedefaultseppunct}\relax
\EndOfBibitem
\bibitem[Berman(2000)]{VRG:berman:2000:NAR}
H.~M. Berman, \emph{Nucleic Acids Res.}, 2000, \textbf{28}, 235--242\relax
\mciteBstWouldAddEndPuncttrue
\mciteSetBstMidEndSepPunct{\mcitedefaultmidpunct}
{\mcitedefaultendpunct}{\mcitedefaultseppunct}\relax
\EndOfBibitem
\bibitem[Neese(2022)]{VRG:neese:2022:WCMS}
F.~Neese, \emph{WIREs Comput Mol Sci}, 2022, \textbf{12}, e1606\relax
\mciteBstWouldAddEndPuncttrue
\mciteSetBstMidEndSepPunct{\mcitedefaultmidpunct}
{\mcitedefaultendpunct}{\mcitedefaultseppunct}\relax
\EndOfBibitem
\bibitem[Liu and Peng(2009)]{VRG:liu:2009:JCP}
W.~Liu and D.~Peng, \emph{J. Chem. Phys.}, 2009, \textbf{131}, 031104\relax
\mciteBstWouldAddEndPuncttrue
\mciteSetBstMidEndSepPunct{\mcitedefaultmidpunct}
{\mcitedefaultendpunct}{\mcitedefaultseppunct}\relax
\EndOfBibitem
\bibitem[Peng \emph{et~al.}(2013)Peng, Middendorf, Weigend, and
  Reiher]{VRG:peng:2013:JCP}
D.~Peng, N.~Middendorf, F.~Weigend and M.~Reiher, \emph{J. Chem. Phys.}, 2013,
  \textbf{138}, 184105\relax
\mciteBstWouldAddEndPuncttrue
\mciteSetBstMidEndSepPunct{\mcitedefaultmidpunct}
{\mcitedefaultendpunct}{\mcitedefaultseppunct}\relax
\EndOfBibitem
\bibitem[Penchoff \emph{et~al.}(2018)Penchoff, Peterson, Quint, Auxier,
  Schweitzer, Jenkins, Harrison, and Hall]{VRG:penchoff:2018:AO}
D.~A. Penchoff, C.~C. Peterson, M.~S. Quint, J.~D. Auxier, G.~K. Schweitzer,
  D.~M. Jenkins, R.~J. Harrison and H.~L. Hall, \emph{ACS Omega}, 2018,
  \textbf{3}, 14127--14143\relax
\mciteBstWouldAddEndPuncttrue
\mciteSetBstMidEndSepPunct{\mcitedefaultmidpunct}
{\mcitedefaultendpunct}{\mcitedefaultseppunct}\relax
\EndOfBibitem
\bibitem[Sun \emph{et~al.}(2020)Sun, Zhang, Banerjee, Bao, Barbry, Blunt,
  Bogdanov, Booth, Chen, Cui, Eriksen, Gao, Guo, Hermann, Hermes, Koh, Koval,
  Lehtola, Li, Liu, Mardirossian, McClain, Motta, Mussard, Pham, Pulkin,
  Purwanto, Robinson, Ronca, Sayfutyarova, Scheurer, Schurkus, Smith, Sun, Sun,
  Upadhyay, Wagner, Wang, White, Whitfield, Williamson, Wouters, Yang, Yu, Zhu,
  Berkelbach, Sharma, Sokolov, and Chan]{VRG:sun:2020:JCP}
Q.~Sun, X.~Zhang, S.~Banerjee, P.~Bao, M.~Barbry, N.~S. Blunt, N.~A. Bogdanov,
  G.~H. Booth, J.~Chen, Z.-H. Cui, J.~J. Eriksen, Y.~Gao, S.~Guo, J.~Hermann,
  M.~R. Hermes, K.~Koh, P.~Koval, S.~Lehtola, Z.~Li, J.~Liu, N.~Mardirossian,
  J.~D. McClain, M.~Motta, B.~Mussard, H.~Q. Pham, A.~Pulkin, W.~Purwanto,
  P.~J. Robinson, E.~Ronca, E.~R. Sayfutyarova, M.~Scheurer, H.~F. Schurkus,
  J.~E.~T. Smith, C.~Sun, S.-N. Sun, S.~Upadhyay, L.~K. Wagner, X.~Wang,
  A.~White, J.~D. Whitfield, M.~J. Williamson, S.~Wouters, J.~Yang, J.~M. Yu,
  T.~Zhu, T.~C. Berkelbach, S.~Sharma, A.~Y. Sokolov and G.~K.-L. Chan,
  \emph{J. Chem. Phys.}, 2020, \textbf{153}, 024109\relax
\mciteBstWouldAddEndPuncttrue
\mciteSetBstMidEndSepPunct{\mcitedefaultmidpunct}
{\mcitedefaultendpunct}{\mcitedefaultseppunct}\relax
\EndOfBibitem
\end{mcitethebibliography}
\bibliographystyle{rsc} %the RSC's .bst file

\end{document}